\documentclass[11pt]{article}

\usepackage[margin=1in]{geometry}
\usepackage[T1]{fontenc}
\usepackage[utf8]{inputenc}
\usepackage{lmodern}
\usepackage{authblk}
\usepackage[numbers,sort&compress]{natbib}
\usepackage{amsmath,amssymb,amsthm,amsfonts}

\usepackage{graphicx,epstopdf}
\usepackage{booktabs}
\usepackage{multirow}
\usepackage{array}
\usepackage{float}
\usepackage[colorlinks=true,linkcolor=blue,citecolor=blue,urlcolor=blue]{hyperref}
\usepackage[numbers]{natbib}

\usepackage{caption}
\usepackage{subcaption}
\usepackage{enumitem}

\theoremstyle{definition}

\theoremstyle{remark}

\title{Dimension Reduction of Higher-Order Dynamical Networks}

\author[1]{Amitosh Tiwari}
\author[2,3]{Chittaranjan Hens}
\author[1]{Prosenjit Kundu}

\affil[1]{CSys Lab, Dhirubhai Ambani University, Gandhinagar, Gujarat 382007, India}
\affil[2]{Center for Computational Natural Sciences and Bioinformatics, International Institute of Information Technology Hyderabad, Gachibowli, Hyderabad 500032, Telangana, India}
\affil[3]{Biomedical Research Center, International Institute of Information Technology Hyderabad, Gachibowli, Hyderabad 500032, Telangana, India}

\begin{document}

\maketitle

\begin{abstract}
Low-dimensional reductions provide a useful framework for studying high-dimensional dynamics on complex networks, but most existing approaches are restricted to pairwise interactions. Here, we develop a one-dimensional reduction for dynamical systems on networks with purely higher-order interactions. The reduction is formulated through an effective higher-order interaction strength ($\beta_{\Delta}$), associated with the triangular interactions of the underlying network and the dynamical system's effective state. We present a theoretical framework for the dimension-reduction approach and validate it across three dynamical models with exclusively higher-order interactions. We find that the reduction accuracy is mainly determined by the homogeneity of node states, i.e., the deviations in state values become very small. Numerical results on synthetic and real networks show that the reduced model captures the effective steady states and transitions of the full system with good accuracy.

\end{abstract}

\section{Introduction}

Resilience refers to the ability of a system to handle disturbances, reorganize itself, and preserve its essential structure and identity \cite{holling1973resilience,walker2004resilience,scheffer2009critical,krakovska2024resilience,schoenmakers2021resilience}. In networked systems, resilience is closely related to the stability of the dynamical processes evolving on the underlying interaction structure \cite{gao2016universal,artime2024robustness,liu2022network}. A loss of resilience can precede sudden regime shifts, often with important consequences for ecological, technological, and socio-economic systems \cite{holling1973resilience,gao2016universal}. Understanding how resilience emerges from the interplay between nonlinear dynamics and network structure, therefore, remains a central problem in complex systems research.

\par
Network science provides a natural framework for studying such questions. A wide range of collective processes in real-world systems have been modeled using dynamical systems on complex networks, including synchronization phenomena \cite{pikovsky2001synchronization,arenas2008physrep,ji2013prl,rodrigues2016kuramoto,kundu2017pre,kundu2018epl,kundu2019chaos,khanra2018pre,khanra2021csf,dutta2025hypergraph,dutta2023perfect,das2025phaselag,dutta2023phase,dutta2024adaptive,dutta2025double,ghosh2025universal}, epidemic spreading and diffusion dynamics \cite{pastor2001prl,granell2013prl,pastor2015rmp,wang2017rpp,mei2017annualreview,colizza2007bmc,higham2021epidemics,yuan2026noise,luo2026temporal}, and many others \cite{albert2002rmp,barrat2008dynamical,newman2010networks,boccaletti2006physrep,dorogovtsev2008rmp}. These studies have shown that the arrangement of interactions can strongly influence the stability, controllability, and resilience of collective dynamics \cite{dawn2026instability,Meena2023stability,Allesina2012}.

\par
A major challenge in analyzing such systems is their high dimensionality. Large networks generally give rise to many coupled nonlinear differential equations, making the direct analysis of equilibrium points, their stability, and their bifurcations difficult. One method for studying these high-dimensional dynamical systems is to reduce their dimension to a lower order, which can predict the collective behavior of the high-dimensional system. Dimension-reduction methods replace the original $N$-dimensional dynamics with one or more effective low-dimensional descriptions \cite{gao2016universal,kundu2022accuracy,jiang2018predicting,laurence2019prx}, thereby making the analysis of steady states, bifurcation points, and tipping phenomena \cite{MaclarenJROS2023} more tractable. The usefulness of this idea has been demonstrated in several settings. A one-dimensional effective equation that describes the global behavior of uncorrelated networks was developed in the seminal work of Gao, Barzel, and Barab'asi \cite{gao2016universal}. Other spectral- and matrix-based methods have also been developed to simplify high-dimensional network dynamics while retaining important structural information \cite{laurence2019prx,thibeault2020threefold,burgio2021compphys}. For pairwise systems, recent work has further examined the conditions under which such reduced descriptions remain accurate \cite{kundu2022accuracy}.

\par
In many real systems, however, interactions are not restricted to pairwise associations. Instead, the state of a node may depend on the joint action of several neighboring nodes. Such group interactions are known as higher-order interactions (HOIs), and they are commonly represented using structures such as hypergraphs and simplicial complexes \cite{battiston2021natphys,boccaletti2023structure,bick2023higher,ji2023physrep,Gracht2024,Majhi2022}. Recent studies have shown that HOIs can qualitatively alter collective behavior. For example, spreading processes on simplicial complexes can exhibit abrupt transitions and bistability that do not arise in standard pairwise models \cite{iacopini2019simplicial,arruda2020contagion}. Higher-order effects can also influence stability and species coexistence in ecological systems \cite{grilli2017higher}, while hypergraph-based epidemic models can display threshold behavior different from that of pairwise descriptions \cite{higham2021epidemics,sun2021hypergraph}.

In contrast, much less is understood when interactions are purely higher-order. For synchronization problems, recent studies have shown that higher-order couplings can generate qualitatively distinct collective behavior, including abrupt desynchronization transitions in pure simplicial complexes \cite{kachhvah2022simplicial}. For other dynamical processes, low-dimensional descriptions have only recently begun to emerge; for example, a one-dimensional reduction has been developed for higher-order contagious phenomena \cite{ghosh2023chaos}.

More broadly, recent work has also reported other higher-order dynamical effects, such as stochastic resonance in oscillator networks with triadic interactions \cite{wang2026network}. These results indicate that higher-order interaction structures can play an important role in the resilience and transition behavior of dynamical systems.

There is still no general framework for reducing networks of a class of dynamical systems \cite{gao2016universal} and identifying when and why a one-dimensional reduction remains accurate in the presence of HOIs. This gap is particularly important for resilience studies, where the usefulness of a reduced model depends not only on its simplicity but also on its ability to capture the effective steady states, bifurcation structures, and transition behavior of networked dynamical systems.   In this study, we represent
a more general reduction technique for systems with purely HOIs by approximating the HOI
interaction using the higher-order (triangular) degree.

\begin{figure}[H]
\centering
    \includegraphics[width=0.75\linewidth]{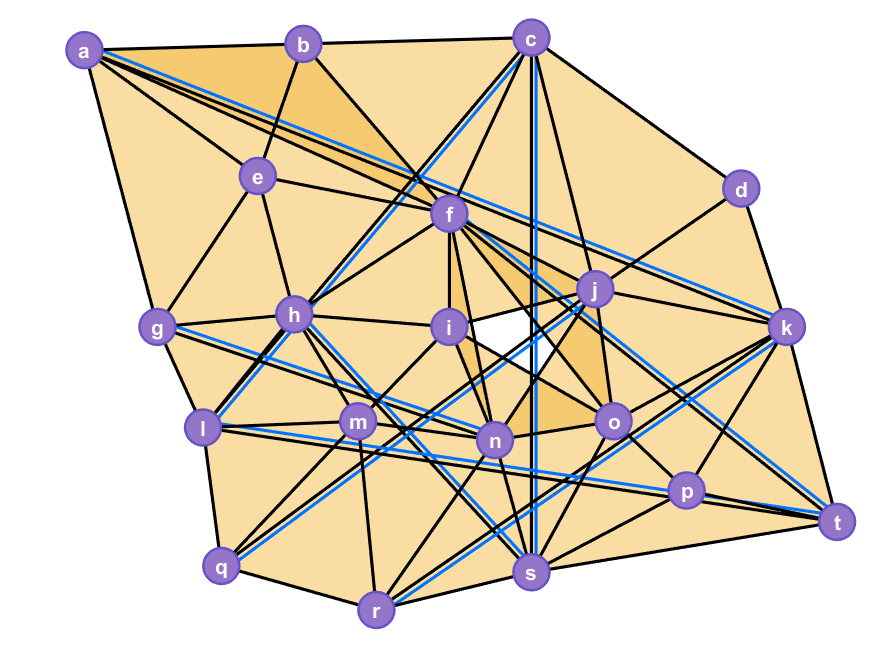}

\caption{Schematic illustration of a network with both pairwise and triangular interactions. The blue edges represent pairwise interactions, the black edges represent triangular-interaction links, and the colored shaded triangular patches indicate the corresponding higher-order triangular motifs.}
\label{fig1}
\end{figure}
In this work, we develop a one-dimensional effective-state reduction framework for dynamical systems on networks with purely higher-order interactions. By introducing the effective higher-order interaction strength parameter $\beta_{\Delta}$ associated with triangular interaction structures, we extend the dimension-reduction idea to purely HOI-driven dynamics. We formulate the reduction through a set of explicit approximations and examine how the validity of these approximations controls the agreement between the full network dynamics and the reduced-order model. Through numerical investigations of three different dynamical systems, namely the double-well potential system \cite{kundu2022rspa}, a nonlinear gene-regulatory system \cite{Meena2023stability,gao2016universal}, and the SIS epidemic model \cite{ghosh2023chaos, Meena2023stability,pastor2015rmp}, with higher-order interactions on synthetic and real networks, we identify the conditions under which the reduced description accurately captures the effective steady states and resilience-related transitions of systems with purely higher-order interactions.

\section{Model}

We consider a system of $N$ nodes whose interactions occur through triangular structures represented by simplicial complexes. The state of node $i$ evolves according to
\begin{equation}
\dot{x}_i = F(x_i) + D_\Delta \sum_{j=1}^{N}\sum_{l=1}^{N} T_{ijl}\, H(x_i,x_j,x_l),
\qquad i=1,\dots,N,
\label{eq:main1}
\end{equation}
where $x_i \in \mathbb{R}$ denotes the dynamical state of node $i$, $F(x_i)$ describes its intrinsic dynamics, and $H(x_i,x_j,x_l)$ represents the combined influence of nodes $j$ and $l$ on node $i$. The tensor $T_{ijl}\in{0,1}$ is the unweighted and undirected third-order interaction tensor, where $T_{ijl}=1$ indicates a higher-order interaction among nodes $i$, $j$, and $l$. The parameter $D_\Delta$ controls the strength of the higher-order coupling.

To derive a one-dimensional reduction of the \(N\)-dimensional dynamics in 
Eq. \eqref{eq:main1}, we first rewrite the local higher-order interaction term as the average contribution of the triangles attached to node \(i\). For this, let
\[
\mathcal{N}_i^\Delta=\{(j,l):T_{ijl}=1\}
\]
denote the set of ordered node pairs that participate in a triangular interaction with node \(i\). Then the higher-order degree of node \(i\) is
\[
k_i^\Delta=\sum_{j=1}^{N}\sum_{l=1}^{N}T_{ijl}
=|\mathcal{N}_i^\Delta|.
\]
Defining
\[
z_{(j,l)}(x_i)=H(x_i,x_j,x_l),
\]
we can write
\[
\sum_{j=1}^{N}\sum_{l=1}^{N}T_{ijl}H(x_i,x_j,x_l)
=
k_i^\Delta
\left\langle z_{(j,l)}(x_i)\right\rangle_{(j,l)\in \mathcal{N}_i^\Delta},
\tag{2.2}
\]
 and
\begin{equation}
\left\langle z_{(j,l)}(x_i)\right\rangle_{(j,l)\in \mathcal{N}_i^\Delta}
=
\frac{1}{k_{\Delta i}}
\sum_{j=1}^{N}\sum_{l=1}^{N}T_{ijl}\,H(x_i,x_j,x_l)
\label{eq:main2b}
\end{equation}
denotes the weighted average higher-order input over the neighbors of node $i$.

If higher-order (triangular) degree correlations are weak, the local neighborhoods of different nodes can be treated as statistically similar. Under this approximation, we replace the neighborhood average in Eq. \eqref{eq:main2b} by
\begin{equation}
\left\langle z_{(j,l)}(x_i)\right\rangle_{(j,l)\in \mathcal{N}_i^\Delta}
\approx
\mathcal{T}(H\!\left(x_i,\mathbf{x},\mathbf{x})\right),
\label{eq:main3}
\end{equation}
where $\mathbf{x}=(x_1,\dots,x_N)^\top$, and $\mathcal{T}$ is the weighted averaging operator defined by
\begin{equation}
\mathcal{T}(\mathbf{z})
=
\frac{\sum_{i=1}^{N} k_{\Delta i} z_i}{\sum_{i=1}^{N} k_{\Delta i}}
=
\frac{\langle k_\Delta z\rangle}{\langle k_\Delta\rangle},
\label{eq:main4}
\end{equation}
for any node-based scalar quantity $\mathbf{z}=(z_1,\dots,z_N)^\top$. Here, $\langle \cdot \rangle$ denotes the unweighted average over all nodes.

Using Eqs. \eqref{eq:main2b}--\eqref{eq:main4}, Eq. \eqref{eq:main1} becomes
\begin{equation}
\frac{dx_i}{dt}
\approx
F(x_i)
+
D_\Delta k_{\Delta i}\,
\mathcal{T}(H\!\left(x_i,\mathbf{x},\mathbf{x}\right)),
\label{eq:main5}
\end{equation}
where, $\mathcal{T}(H\!\left(x_i,\mathbf{x},\mathbf{x}\right))$ represents the average input to node $i$ from the neighboring pair  of nodes $(j,l$) with the weight of its triangular degree. Now, with an approximation $\mathcal{T}(H\!\left(x_i,\mathbf{x},\mathbf{x}\right))\approx H\!\left(x_i,\mathcal{T}(\mathbf{x}),\mathcal{T}(\mathbf{x}\right))$ we can rewrite the Eq. \eqref{eq:main5} in vector form as
\begin{equation}
\frac{d\mathbf{x}}{dt}
\approx
F(\mathbf{x})
+
D_\Delta\,\mathbf{k}_{\Delta}\circ
H\!\left(\mathbf{x},\mathcal{T}(\mathbf{x}),\mathcal{T}(\mathbf{x})\right),
\label{eq:main6}
\end{equation}
where
\[
H\!\left(\mathbf{x},\mathcal{T}(\mathbf{x}),\mathcal{T}(\mathbf{x})\right)
=
\big(
H(x_1,\mathcal{T}(\mathbf{x}),\mathcal{T}(\mathbf{x})),
\dots,
H(x_N,\mathcal{T}(\mathbf{x}),\mathcal{T}(\mathbf{x}))
\big)^\top,
\]
\[
\mathbf{k}_{\Delta}=(k_{\Delta 1},\dots,k_{\Delta N})^\top,
\]
Moreover, $\circ$ denotes the Hadamard product.

Since $\mathcal{T}$ is a linear operator, applying it to Eq. \eqref{eq:main6} gives
\begin{align}
\frac{d\,\mathcal{T}(\mathbf{x})}{dt}
&=
\mathcal{T}\!\left(F(\mathbf{x})\right)
+
D_\Delta\,
\mathcal{T}\!\left(
\mathbf{k}_{\Delta}\circ
H\!\left(\mathbf{x},\mathcal{T}(\mathbf{x}),\mathcal{T}(\mathbf{x})\right)
\right)
\nonumber\\
&\approx
F\!\left(\mathcal{T}(\mathbf{x})\right)
+
D_\Delta\,
\mathcal{T}(\mathbf{k}_{\Delta})\,
H\!\left(\mathcal{T}(\mathbf{x}),\mathcal{T}(\mathbf{x}),\mathcal{T}(\mathbf{x})\right).
\label{eq:main7}
\end{align}
To obtain the last line in Eq. \eqref{eq:main7}, we use the approximations
\[
\mathcal{T}(F(\mathbf{x})) \approx F(\mathcal{T}(\mathbf{x}))
\]
and
\[
\mathcal{T}\!\left(
\mathbf{k}_{\Delta}\circ
H\!\left(\mathbf{x},\mathcal{T}(\mathbf{x}),\mathcal{T}(\mathbf{x})\right)
\right)
\approx
\mathcal{T}(\mathbf{k}_{\Delta})\,
H\!\left(\mathcal{T}(\mathbf{x}),\mathcal{T}(\mathbf{x}),\mathcal{T}(\mathbf{x})\right).
\]

Following Gao \textit{et al.} \cite{gao2016universal}, we first define the effective state as
\begin{equation}
x = x_{\mathrm{eff}}^{\mathrm{HOI}}
=
\mathcal{T}(\mathbf{x})
=
\frac{\langle k_\Delta x\rangle}{\langle k_\Delta\rangle},
\label{eq:main8}
\end{equation}
which is the weighted average of $x_i$'s weighted by the triangular degree. We also define
\begin{equation}
\beta_{\Delta}=\beta_{\mathrm{eff}}^{\mathrm{HOI}}
=
\mathcal{T}(\mathbf{k}_{\Delta})
=
\frac{\langle k_\Delta^2\rangle}{\langle k_\Delta\rangle}.
\label{eq:main9}
\end{equation}

Substituting $\mathcal{T}(\mathbf{x})=x_{\mathrm{eff}}^{\mathrm{HOI}}=x$ and $\mathcal{T}(\mathbf{k}_{\Delta})=\beta_{\Delta}$ into Eq. \eqref{eq:main7}, we obtain the one-dimensional reduced dynamics
\begin{equation}
\frac{dx}{dt}
=
F(x)
+
D_{\Delta}\beta_{\Delta}H(x,x,x).
\label{eq:main11}
\end{equation}

Under the higher-order mean-field closure, the reduced dynamics in Eq. \eqref{eq:main11} are approximately satisfied at $(x,\beta_\Delta)=(x_{\mathrm{eff}}^{\mathrm{HOI}},\beta_\Delta)$. When this closure captures the collective behavior well, the equilibria and stability properties of the reduced system can be used to predict the resilience of the full $N$-dimensional higher-order dynamical system.

The stable equilibria of the reduced dynamics are obtained by setting the right-hand side of Eq. \eqref{eq:main11} to zero and retaining only the linearly stable solutions. We denote the corresponding stable branch by $x^*(\beta_\Delta)$. Provided that higher-order structural heterogeneity does not induce localized collective modes, this branch approximates the effective equilibrium state of the original system, denoted by $x_{\mathrm{eff}}^{\mathrm{HOI}^*}$.

For Eq. \eqref{eq:main11} to accurately approximate the dynamics of $x_{\mathrm{eff}}$, the following approximations must hold:
\begin{align}
\text{Approximation }(A_2):\quad
&\mathcal{T}(F(\mathbf{x}))
\approx
F(\mathcal{T}(\mathbf{x})),
\label{eq:main12}
\\[4pt]
\text{Approximation }(A_3):\quad
& \mathcal{T}(H(x_i,\mathbf{x},\mathbf{x}))
\approx
H\!\left(x_i,\mathcal{T}(\mathbf{x}),\mathcal{T}(\mathbf{x})\right),
\label{eq:main13}
\\[4pt]
\text{Approximation }(A_4):\quad
&\mathcal{T}\!\left(
\mathbf{k}_{\Delta}\circ
H\!\left(\mathbf{x},\mathcal{T}(\mathbf{x}),\mathcal{T}(\mathbf{x})\right)
\right)
\approx
\mathcal{T}(\mathbf{k}_{\Delta})\,
H\!\left(\mathcal{T}(\mathbf{x}),\mathcal{T}(\mathbf{x}),\mathcal{T}(\mathbf{x})\right).
\label{eq:main14}
\end{align}

To understand when the higher-order reduction is accurate, we examine the validity of approximations $(A_2)$, $(A_3)$, and $(A_4)$ for different dynamical systems.The approximation conditions \(A_2\), \(A_3\), and \(A_4\) help identify when the reduced higher-order model is expected to be accurate. Condition \(A_2\) checks whether the intrinsic node dynamics are compatible with the averaging map, \(A_3\) checks whether the interaction term can be written in terms of the effective state, and \(A_4\) checks whether the higher-order structural term can also be closed at the effective-state level. Therefore, their validity depends on the node dynamics, the coupling function, and the higher-order interaction structure. For example, approximation \(A_4\) does not hold for diffusive higher-order coupling. In that case, the right-hand side of Eq. \eqref{eq:main14} vanishes because
\[
H(\mathcal{T}(\mathbf{x}),\mathcal{T}(\mathbf{x}),\mathcal{T}(\mathbf{x}))=0
\]
when all arguments are equal, whereas the left-hand side is generally nonzero since \(x_i\) need not be equal to \(\mathcal{T}(\mathbf{x})\) for all nodes. This shows that diffusive higher-order coupling is an important boundary case where the reduction may become less accurate.

\section{Approximation Validation :}
To investigate the accuracy of approximations (A$_{2}$), (A$_{3}$), and (A$_{4}$), we calculate the ratio of the right-hand side to the left-hand side of Eqs. ~\eqref{eq:main12},~\eqref{eq:main13} and Eq. ~\eqref{eq:main14}, which we refer to as :
\begin{equation}
R_2 = \frac{\mathcal{T}(F(\mathbf{x}))}{F(\mathcal{T}(\mathbf{x}))}
\end{equation}

\begin{equation}
R_3 = \frac{\mathcal{T}\!\left(H(x_i,\mathbf{x},\mathbf{x})\right)}
{H\!\left(x_i,\mathcal{T}(\mathbf{x}),\mathcal{T}(\mathbf{x})\right)}
\end{equation}

\begin{equation}
R_4 =
\frac{\mathcal{T}\!\left(k_{\Delta} \circ H(\mathbf{x},\mathcal{T}(\mathbf{x}),\mathcal{T}(\mathbf{x}))\right)}
{\mathcal{T}(k_{\Delta})\,H\!\left(\mathcal{T}(\mathbf{x}),\mathcal{T}(\mathbf{x}),\mathcal{T}(\mathbf{x})\right)}
\end{equation}
\subsection*{Double-well }

We consider
\begin{equation}
\frac{dx_i}{dt}
=
-(x_i-r_1)(x_i-r_2)(x_i-r_3)
+
D_{\Delta}\sum_{j=1}^{N}\sum_{l=1}^{N}T_{ijl}\,x_jx_l,
\label{eq:DW_full}
\end{equation}
where $x_i$ is the state of the $i$th node and $D_{\Delta}$ is the higher-order coupling strength. In the absence of coupling, the single-node dynamics have two stable equilibria at $x=r_1$ and $x=r_3$, with $r_1<r_2<r_3$.

The corresponding reduced one-dimensional dynamics is given by
\begin{equation}
\frac{dx}{dt}
=
-(x-r_1)(x-r_2)(x-r_3)
+
D_{\Delta}\beta_{\Delta}x^2.
\label{eq:DW_red}
\end{equation}

\textbf{Approximation $A_2$}

For the double-well system,
\[
F(x)=-(x-r_1)(x-r_2)(x-r_3).
\]
Expanding this term gives
\[
F(x)
=
-x^3
+
(r_1+r_2+r_3)x^2
-
(r_1r_2+r_2r_3+r_3r_1)x
+
r_1r_2r_3.
\]
Using
\[
x_{\mathrm{eff}}^{\mathrm{HOI}}=\mathcal{T}(\mathbf{x})
=
\frac{\langle k_{\Delta}x\rangle}{\langle k_{\Delta}\rangle},
\]
we obtain
\begin{equation}
R_2
=
\frac{
-\dfrac{\langle k_{\Delta}x^3\rangle}{\langle k_{\Delta}\rangle}
+
(r_1+r_2+r_3)\dfrac{\langle k_{\Delta}x^2\rangle}{\langle k_{\Delta}\rangle}
-
(r_1r_2+r_2r_3+r_3r_1)x_{\mathrm{eff}}^{\mathrm{HOI}}
+
r_1r_2r_3
}{
-x_{\mathrm{eff}}^{\mathrm{HOI}^3}
+
(r_1+r_2+r_3)x_{\mathrm{eff}}^{\mathrm{HOI}^2}
-
(r_1r_2+r_2r_3+r_3r_1)x_{\mathrm{eff}}^{\mathrm{HOI}}
+
r_1r_2r_3
}.
\label{eq:DW_R2}
\end{equation}
For a homogeneous network, a simple sufficient condition for $R_2\sim 1$ is that the deviation of  $x$ becomes sufficiently small.

\textbf{Approximation $A_3$}

For the higher-order interaction term,
\[
H(x_i,x_j,x_l)=x_jx_l.
\]
Therefore,
\[
H\!\left(x_i,\mathcal{T}(\mathbf{x}),\mathcal{T}(\mathbf{x})\right)
=
x_{\mathrm{eff}}^{\mathrm{HOI}^2}.
\]
Hence
\begin{equation}
R_3
=
\frac{\mathcal{T}(x_jx_l)}{x_{\mathrm{eff}}^{\mathrm{HOI}^2}}.
\label{eq:DW_R3_step1}
\end{equation}
Using the operator definition,
\begin{equation}
\mathcal{T}(x_jx_l)
=
\frac{\langle k_{\Delta}x_jx_l\rangle}{\langle k_{\Delta}\rangle}.
\label{eq:DW_R3_step2}
\end{equation}
Thus
\begin{equation}
R_3
=
\frac{\langle k_{\Delta}x_jx_l\rangle}
{\langle k_{\Delta}\rangle x_{\mathrm{eff}}^{\mathrm{HOI}^2}}.
\label{eq:DW_R3}
\end{equation}
For a homogeneous network $R_3 \sim 1$ is expected if the deviation of $x$ is close to zero.

\textbf{Approximation $A_4$}

For the double-well coupling,
\[
H\!\left(\mathbf{x},\mathcal{T}(\mathbf{x}),\mathcal{T}(\mathbf{x})\right)
=
x_{\mathrm{eff}}^{\mathrm{HOI}^2},
\]
since the higher-order term depends only on the second and third arguments. Therefore,
\begin{equation}
R_4
=
\frac{
\mathcal{T}\!\left(
\mathbf{k}_{\Delta}\circ x_{\mathrm{eff}}^{\mathrm{HOI}^2}
\right)
}{
\mathcal{T}(\mathbf{k}_{\Delta})x_{\mathrm{eff}}^{\mathrm{HOI}^2}
}.
\label{eq:DW_R4_step1}
\end{equation}
Because $x_{\mathrm{eff}}^{\mathrm{HOI}^2}$ is a scalar, it cancels from the numerator and denominator, yielding
\begin{equation}
R_4=1.
\label{eq:DW_R4}
\end{equation}
Hence the approximation $(A_4)$ is trivial for the double-well potential system with higher-order coupling considered here.

\subsection*{Gene Regulatory System}

We consider a model of a gene regulatory system governed by the Michaelis-Menten equation:
\begin{equation}
\frac{dx_i}{dt}
=
- B x_i^{f}
+
D_{\Delta}\sum_{j=1}^{N}\sum_{l=1}^{N}
T_{ijl}
\frac{(x_jx_l)^{h}}{1 +(x_jx_l)^{h}},
\label{eq:gene_full}
\end{equation}
where $x_i$ represents the expression level of gene $i$. The first term describes degradation, and the second term represents higher-order activation.

The corresponding reduced one-dimensional dynamics are
\begin{equation}
\frac{dx}{dt}
=
- B x^{f}
+
D_{\Delta}\beta_{\Delta}
\frac{x^{2h}}{1 + x^{2h}}.
\label{eq:gene_red}
\end{equation}

\textbf{Approximation $A_2$}

For the gene-regulatory system,
\[
F(x_i)=-Bx_i^{f}, 
\]
Using
\[
x_{\mathrm{eff}}=\mathcal{T}(\mathbf{x})
=
\frac{\langle k_{\Delta}x\rangle}{\langle k_{\Delta}\rangle},
\]
we obtain
\begin{equation}
R_2
=
\frac{\mathcal{T}(F(\mathbf{x}))}{F(\mathcal{T}(\mathbf{x}))}
=
\frac{-B\,\mathcal{T}(x^{f})}{-B\,x_{\mathrm{eff}}^{\mathrm{HOI}^f}}
=
\frac{\langle k_{\Delta}x^{f}\rangle}
{\langle k_{\Delta}\rangle x_{\mathrm{eff}}^{\mathrm{HOI}^f}}.
\label{eq:gene_R2}
\end{equation}
For a homogeneous network, a simple sufficient condition for $R_2\sim 1$ is that the deviation of  $x$ becomes sufficiently small. In particular, we have considered that $f=1$ makes this approximation trivial.

\textbf{Approximation $A_3$}

For the higher-order interaction term,
\[
H(x_i,x_j,x_l)
=
\frac{(x_jx_l)^{h}}{1+(x_jx_l)^{h}}.
\]
Therefore,
\[
H\!\left(x_i,\mathcal{T}(\mathbf{x}),\mathcal{T}(\mathbf{x})\right)
=
\frac{x_{\mathrm{eff}}^{2h}}{1+x_{\mathrm{eff}}^{\mathrm{HOI}^{2h}}}.
\]
Hence
\begin{equation}
R_3
=
\frac{
\mathcal{T}\!\left(
\frac{(x_jx_l)^{h}}{1+(x_jx_l)^{h}}
\right)
}{
\frac{x_{\mathrm{eff}}^{\mathrm{HOI}^{2h}}}{1+x_{\mathrm{eff}}^{\mathrm{HOI}^{2h}}}
}.
\label{eq:gene_R3_step1}
\end{equation}
Using the operator definition,
\begin{equation}
\mathcal{T}\!\left(
\frac{(x_jx_l)^{h}}{1+(x_jx_l)^{h}}
\right)
=
\frac{
\left\langle
k_{\Delta}\,
\frac{(x_jx_l)^{h}}{1+(x_jx_l)^{h}}
\right\rangle
}{
\langle k_{\Delta}\rangle
}.
\label{eq:gene_R3_step2}
\end{equation}
Thus
\begin{equation}
R_3
=
\frac{
\left\langle
k_{\Delta}\,
\frac{(x_jx_l)^{h}}{1+(x_jx_l)^{h}}
\right\rangle
}{
\langle k_{\Delta}\rangle\,
\frac{x_{\mathrm{eff}}^{2h}}{1+x_{\mathrm{eff}}^{2h}}
}.
\label{eq:gene_R3}
\end{equation}
A simple sufficient condition for $R_3 \sim 1$ is  that the states of all the nodes are tightly packed, i.e, $std(x)\sim 0$.

\textbf{Approximation $A_4$}

For the gene regulatory system,
\[
H\!\left(\mathbf{x},\mathcal{T}(\mathbf{x}),\mathcal{T}(\mathbf{x})\right)
=
\frac{x_{\mathrm{eff}}^{\mathrm{HOI}^{2h}}}{1+x_{\mathrm{eff}}^{\mathrm{HOI}^{2h}}},
\]
since the higher-order term depends only on the second and third arguments. Therefore,
\begin{equation}
R_4
=
\frac{
\mathcal{T}\!\left(
\mathbf{k}_{\Delta}\circ
\frac{x_{\mathrm{eff}}^{\mathrm{HOI}^{2h}}}{1+x_{\mathrm{eff}}^{\mathrm{HOI}^{2h}}}
\right)
}{
\mathcal{T}(\mathbf{k}_{\Delta})\,
\frac{x_{\mathrm{eff}}^{\mathrm{HOI}^{2h}}}{1+x_{\mathrm{eff}}^{\mathrm{HOI}^{2h}}}
}.
\label{eq:gene_R4_step1}
\end{equation}
Because
\[
\frac{x_{\mathrm{eff}}^{\mathrm{HOI}^{2h}}}{1+x_{\mathrm{eff}}^{\mathrm{HOI}^{2h}}}
\]
is a scalar, it cancels from the numerator and denominator, yielding
\begin{equation}
R_4=1.
\label{eq:gene_R4}
\end{equation}
Hence the approximation $(A_4)$ trivially holds for the gene regulatory higher-order coupling considered here.
\subsection*{SIS Model}

We consider the deterministic approximation to the susceptible-infectious-susceptible (SIS) dynamics:
\begin{equation}
\dot{x}_i
=
-\mu x_i
+
\gamma (1-x_i)\sum_{j=1}^{N}\sum_{l=1}^{N}T_{ijl}(x_jx_l),
\label{eq:sis_full}
\end{equation}
where $\mu$ is the recovery rate and $\gamma$ is the higher-order infection rate.

The corresponding reduced one-dimensional dynamics are
\begin{equation}
\dot{x}
=
-\mu x
+
\gamma \beta_{\Delta}(1-x)x^2.
\label{eq:sis_red}
\end{equation}

\textbf{Approximation $A_2$}

For the SIS model,
\[
F(x_i)=-\mu x_i.
\]

is a linear function and hence
$R_2=1$
Hence approximation $(A_2)$ is trivial.

\textbf{Approximation $A_3$}

For the higher-order infection term,
\[
H(x_i,x_j,x_l)=(1-x_i)x_jx_l.
\]
Therefore,
\[
H\!\left(x_i,\mathcal{T}(\mathbf{x}),\mathcal{T}(\mathbf{x})\right)
=
(1-x_i)x_{\mathrm{eff}}^{\mathrm{HOI}^2}.
\]
Hence
\begin{equation}
R_3
=
\frac{
\mathcal{T}\!\left((1-x_i)x_jx_l\right)
}{
(1-x_i)x_{\mathrm{eff}}^{\mathrm{HOI}^2}
}.
\label{eq:sis_R3_step1}
\end{equation}
Using the operator definition,
\begin{equation}
\mathcal{T}\!\left((1-x_i)x_jx_l\right)
=
\frac{\left\langle k_{\Delta}(1-x_i)x_jx_l\right\rangle}{\langle k_{\Delta}\rangle}.
\label{eq:sis_R3_step2}
\end{equation}
Thus
\begin{equation}
R_3
=
\frac{
\left\langle k_{\Delta}(1-x_i)x_jx_l\right\rangle
}{
\langle k_{\Delta}\rangle (1-x_i)x_{\mathrm{eff}}^{\mathrm{HOI}^2}
}.
\label{eq:sis_R3}
\end{equation}

A sufficient condition for$R_3 \sim 1$
In a homogeneous network, the deviation of 

$x$ remains sufficiently small.

\textbf{Approximation $A_4$}

For the SIS model,
\[
H\!\left(\mathbf{x},\mathcal{T}(\mathbf{x}),\mathcal{T}(\mathbf{x})\right)
=
(1-\mathbf{x})x_{\mathrm{eff}}^{\mathrm{HOI}^2}.
\]
Therefore,
\begin{equation}
R_4
=
\frac{
\mathcal{T}\!\left(
\mathbf{k}_{\Delta}\circ(1-\mathbf{x})x_{\mathrm{eff}}^{\mathrm{HOI}^2}
\right)
}{
\mathcal{T}(\mathbf{k}_{\Delta})(1-x_{\mathrm{eff}}^{\mathrm{HOI}^2})x_{\mathrm{eff}}^{\mathrm{HOI}^2}
}.
\label{eq:sis_R4_step1}
\end{equation}
Because $x_{\mathrm{eff}}^{\mathrm{HOI}^2}$ is a scalar, it cancels from the numerator and denominator, and we obtain
\begin{equation}
R_4
=
\frac{
\mathcal{T}\!\left(
\mathbf{k}_{\Delta}\circ(1-\mathbf{x})
\right)
}{
\mathcal{T}(\mathbf{k}_{\Delta})(1-x_{\mathrm{eff}}^{\mathrm{HOI}^2})
}.
\label{eq:sis_R4_step2}
\end{equation}
Using the definition of $\mathcal{T}$, this becomes
\begin{equation}
R_4
=
\frac{
\langle k_{\Delta}^2(1-\mathbf{x})\rangle
}{
\langle k_{\Delta}^2\rangle(1-x_{\mathrm{eff}}^{\mathrm{HOI}^2})
}.
\label{eq:sis_R4}
\end{equation}
Equivalently,
\begin{equation}
R_4
=
\frac{
\langle k_{\Delta}^2\rangle-\langle k_{\Delta}^2\mathbf{x}\rangle
}{
\langle k_{\Delta}^2\rangle(1-x_{\mathrm{eff}}^{\mathrm{HOI}^2})
}.
\label{eq:sis_R4_alt}
\end{equation}
Thus approximation $(A_4)$ is accurate when
\[
\langle k_{\Delta}^2\mathbf{x}\rangle
\approx
\langle k_{\Delta}^2\rangle x_{\mathrm{eff}}^{\mathrm{HOI}^2}.
\]

which holds  for deviation $\sim 0$.
%\begin{table}[h]
%\centering
%\caption{Conditions for $R_2=1$, $R_3=1$, and %$R_4=1$ for the three dynamical systems.}
%\label{tab:ratio_conditions}
%\renewcommand{\arraystretch}{1.3}
%\small
%\begin{tabular}{|p{2.4cm}|p{2.8cm}|p{2.8cm}|p{2.8cm}|}
%\hline
%System & $R_2=1$ when & $R_3=1$ when & $R_4=1$ when \\
%\hline
%Double well 
%& $x_i=x_{\mathrm{eff}}$ for all $i$
%& $x_j=x_l=x_{\mathrm{eff}}$
%& exact \\

%Gene regulatory 
%& $x_i=x_{\mathrm{eff}}$ for all $i$
%& $x_j=x_l=x_{\mathrm{eff}}$
%& exact
 %\\

%SIS 
%& exact
%& $x_i=x_j=x_l=x_{\mathrm{eff}}$
%& $\langle k_\Delta^2 x\rangle=\langle k_\Delta^2\rangle x_{\mathrm{eff}}$
% \\
%\hline
%\end{tabular}
%\end{table}
\section{Numerical Validation}

For numerical validation, we test the proposed reduction on both synthetic and real networks. For the synthetic networks, we use Erd\H{o}s--R'enyi (ER) random networks and Barab'asi--Albert (BA) scale-free networks, which represent  two distinct network structures. To study the effect of structural changes, nodes are removed stepwise. At each step, 1\% of the total nodes are removed, and all relevant quantities are recalculated. After each removal step, we consider only the largest connected component (LCC) of the remaining network to ensure network connectedness. This process continues until only 1\% of the original nodes remain or the remaining nodes become isolated. We apply the same method to real-world networks to check the robustness of the results. We use two initial conditions: a high and a low one. To study the effect of node dynamics, we consider three different dynamical systems with higher-order interactions, as described below.

\subsection{Double-well system}

We first test the reduction for the double-well system with purely higher-order interactions:
\begin{equation}
\frac{dx}{dt}
=
-(x-r_1)(x-r_2)(x-r_3)
+
D_{\Delta}\beta_{\Delta}x^2.
\end{equation}
In the uncoupled case, the local dynamics have two stable states separated by one unstable state. Here, $x$ denotes the effective steady-state value, $r_1$, $r_2$, and $r_3$ set the local double-well structure, $D_{\Delta}$ is the higher-order coupling strength, and $\beta_{\Delta}$ is the effective higher-order interaction strength. Throughout this subsection, we use
$r_1=1$, $r_2=2$, $r_3=5$, and $D_{\Delta}=0.01$.
For numerical validation, we integrate the networked system with low and high initial conditions until a steady state is reached. To probe structural degradation, nodes are removed in steps of $1\%$. After each step, we retain the largest connected component (LCC), recompute $\beta_{\Delta}$, and evaluate the new effective steady state.

\begin{figure}
    \centering
    \includegraphics[width=1\linewidth]{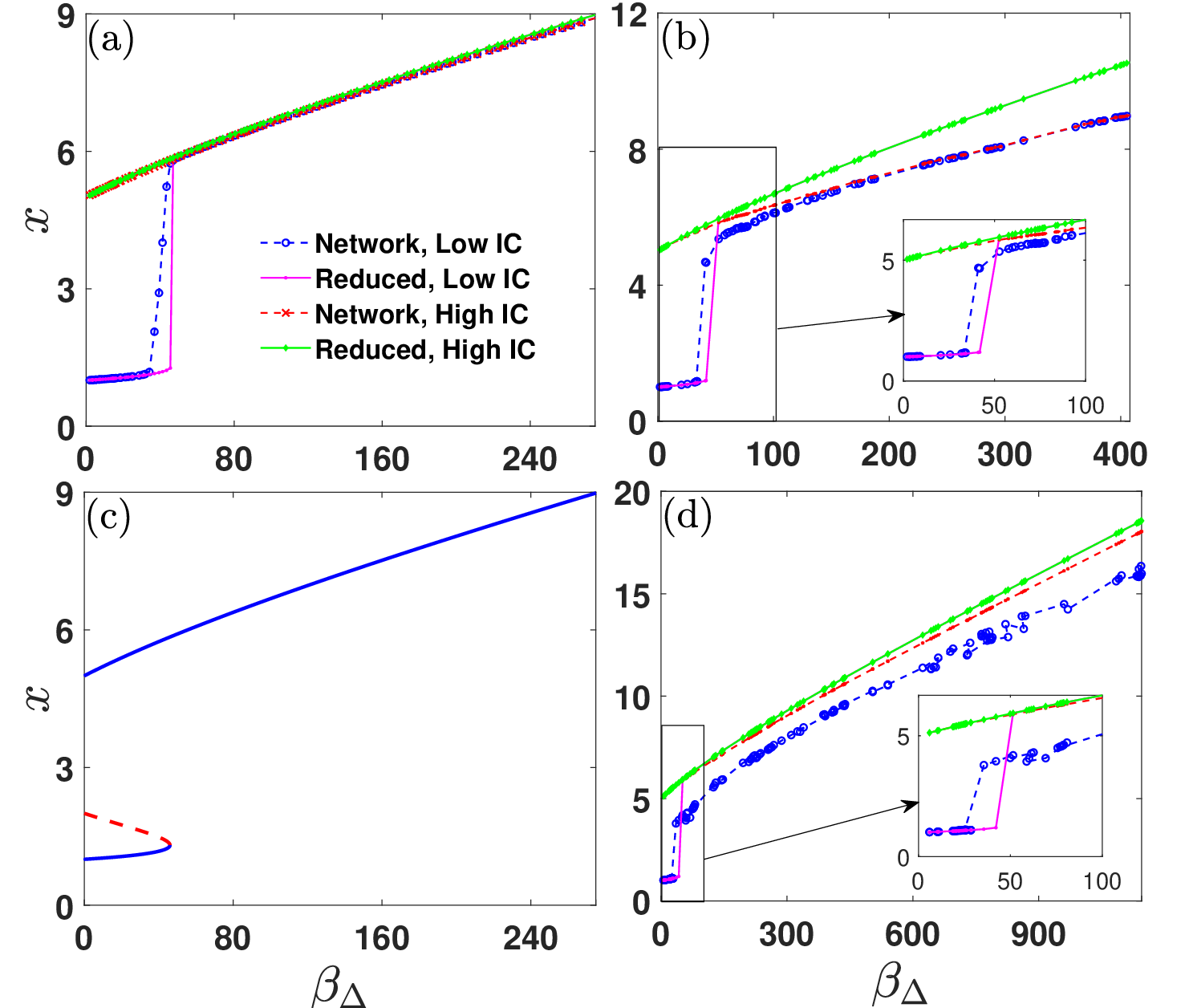}
    \caption{\textbf{Double-well system with higher-order interactions.} Steady-state value $x$ as a function of the effective higher-order interaction strength $\beta_{\Delta}$. Panels (a) and (b) compare the full-network simulations with the one-dimensional reduction for Erd\H{o}s--R\'enyi and Barab\'asi--Albert networks, with $p=0.1$ and $m=12$, respectively, where $m$ is the number of edges attached by each new node during growth. Panel (c) shows the bifurcation diagram of the reduced system, including the stable and unstable branches. Panel (d) shows the corresponding comparison for the GR-QC collaboration network. The reduction captures the main branch structure and the transition from the low state to the high state, with stronger agreement for the synthetic networks than for the real-network low-initial-condition branch.}

\label{fig:validation_DW}
\end{figure}

For the Erd\H{o}s--R\'enyi network, we use $N=500$, $p=0.1$, and remove $5$ randomly selected nodes at each step.The comparison in Fig.~\ref{fig:validation_DW}(a) shows that the reduced model closely follows the full-network branches for both $0.01$ and $5.5$ initial conditions. The agreement is especially good away from the transition region, and the jump from the low branch to the high branch is captured at nearly the same $\beta_{\Delta}$.
For the Barab\'asi--Albert network, we again take $N=500$, with $m=12$, where $m$ is the number of edges attached by each new node during growth. Fig.~\ref{fig:validation_DW}(b) shows that the reduced model still reproduces the main branch structure and the transition. However, because of the higher heterogeneity, the deviation is larger than in the ER case, particularly along the low-initial-condition branch at larger $\beta_{\Delta}$.
Fig.~\ref{fig:validation_DW}(c) shows the bifurcation diagram of the reduced one-dimensional system. The low stable branch, high stable branch, and intermediate unstable branch are clearly separated, confirming that the reduced dynamics preserve the bistable structure of the double-well system.
\\ For the real-network test, we use the GR-QC collaboration network with $5242$ nodes from the \href{https://snap.stanford.edu/data/ca-GrQc.html}{SNAP ca-GrQc dataset}. Here, about $52$ nodes are removed at each step.
As shown in Fig.~\ref{fig:validation_DW}(d), the reduced model captures the overall increase of the effective steady state and the transition from the low-state branch to the high-state branch. However, the agreement is weaker than in the synthetic networks. This discrepancy can be understood in terms of the higher-order organization of the real collaboration network. In contrast to synthetic networks that we consider, where the coefficient of assortativity is low, this particular real network have a very high coefficient of assortativity ($\sim 0.4$) and also it's highly heterogeneous structure leads to some discrepancy when the effective state is compared with the reduced model.
%triangular interactions are generated in a more controlled way, real networks can exhibit a more heterogeneous distribution of triangle participation across nodes. As a result, some nodes may experience much stronger local higher-order feedback than others. Such localized variation in triangular input is difficult to represent completely by a single effective interaction strength \(\beta_\Delta\). The effect is more visible on the low-initial-condition branch, where local differences in higher-order feedback can produce relatively larger deviations from the effective state. Thus, while the reduced model remains informative at the level of the dominant effective-state trend, Fig.~\ref{fig:validation_DW}(d) also illustrates the expected loss of accuracy when the higher-order structure of the real network is strongly heterogeneous.
The approximation-level behavior is shown in Appendix Figs. \ref{DW_ER} - \ref{DW_REAL}. For the ER network Fig. \ref{DW_ER}, the closures remain close to the ideal relations over most of the parameter range, and the relative error is small except for a narrow peak near the transition. For the BA network Fig. \ref{DW_BA}, the same pattern is observed, but with a larger spread in the intrinsic-term closure and a broader low-IC error. For the real network Fig. \ref{DW_REAL}, the approximation quality is mixed: the factorization step remains accurate, while the intrinsic and interaction closures show visible deviations from the ideal line. This explains why the real-network fit is less accurate, while still supporting the use of the reduced model to capture the main branch structure and transition behavior.\\

\subsection{Gene regulatory system}

We next test the reduction for the gene regulatory system with purely higher-order interactions:
\begin{equation}
\frac{dx}{dt}
=
- Bx^{f}
+
D_{\Delta}\beta_{\Delta}\frac{x^{2h}}{1+x^{2h}}.
\end{equation}
Here, $x$ denotes the effective steady-state gene expression level. The first term represents degradation, with $B$ as the degradation strength and $f$ controlling its nonlinearity. The second term represents higher-order activation, with $D_{\Delta}$ as the higher-order coupling strength, $h$ as the Hill exponent, and $\beta_{\Delta}$ as the effective higher-order interaction strength. Throughout this paper, we use
\[
B=1,\qquad f=1,\qquad D_{\Delta}=0.05,\qquad h=2.
\]

The steady states satisfy
\begin{equation}
-Bx^{f}
+
D_{\Delta}\beta_{\Delta}\frac{x^{2h}}{1+x^{2h}}
=0.
\end{equation}
Thus, $x=0$ is always a fixed point. For $x\neq 0$, the nonzero steady states satisfy
\begin{equation}
x^{\,f-2h}+x^{\,f}
=
\frac{D_{\Delta}\beta_{\Delta}}{B}.
\end{equation}
With the parameter values used here, this becomes
\begin{equation}
x^{-3}+x
=
0.05\,\beta_{\Delta}.
\end{equation}

\begin{figure}
    \centering
    \includegraphics[width=1\linewidth]{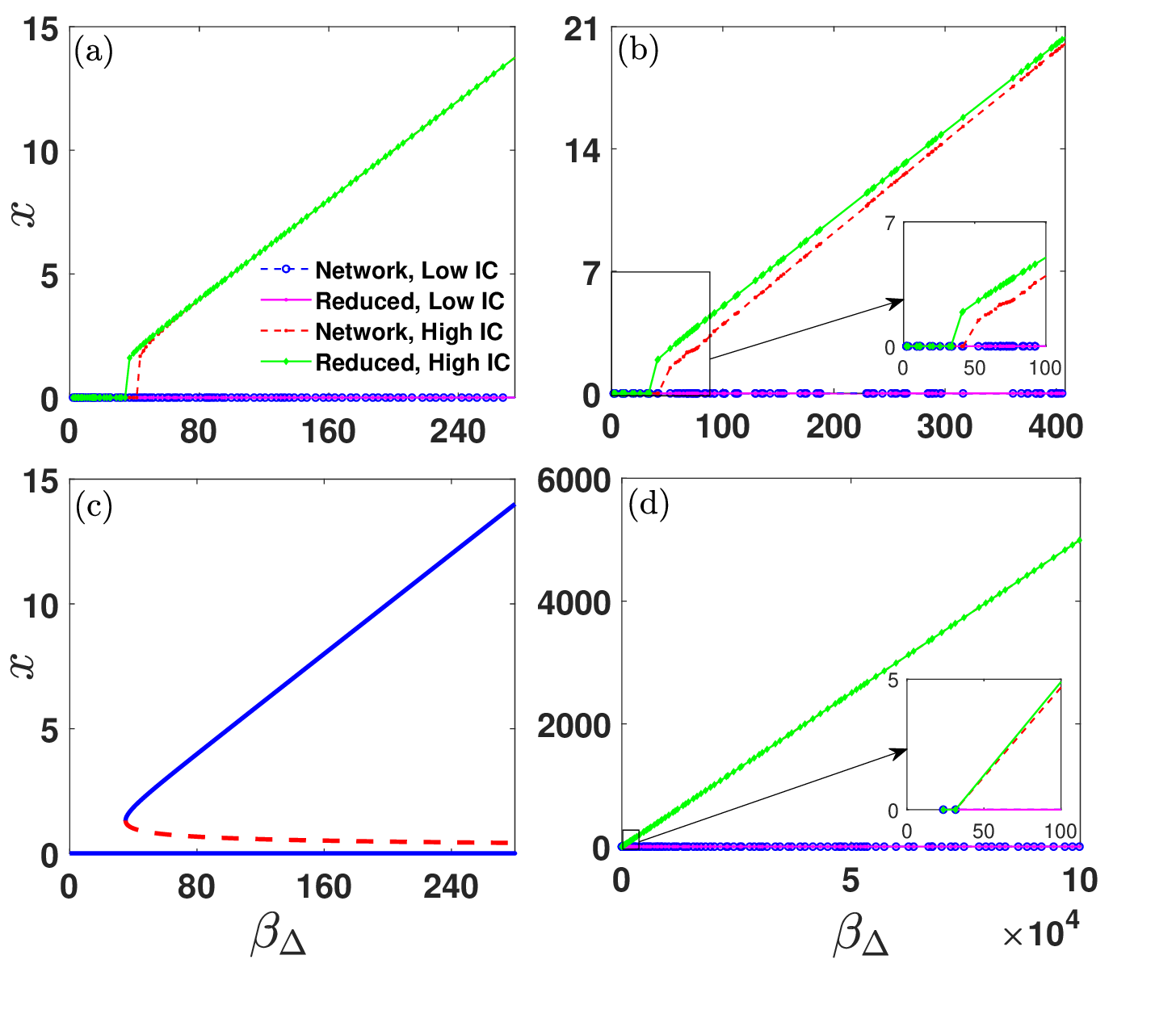}
  \caption{\textbf{Gene regulatory system with higher-order interactions.} Steady-state value $x$ as a function of the effective higher-order interaction strength $\beta_{\Delta}$. Panels (a) and (b) compare the full-network simulations with the one-dimensional reduction for Erd\H{o}s--R\'enyi and Barab\'asi--Albert networks, with $p=0.1$ and $m=12$, respectively, where $m$ is the number of edges attached by each new node during growth. Panel (c) shows the bifurcation diagram of the reduced system, including the stable and unstable branches. Panel (d) shows the corresponding comparison for the \textit{C.~elegans} integrated functional interaction network from the CCSB Interactome Database. The reduction reproduces the main branch structure and the transition from the low-expression state to the high-expression state, with strong agreement for both the synthetic and real networks.}
\label{fig:validation_gene}
\end{figure}

For the full network dynamics, we integrate with low and high initial conditions  until a steady state is reached. To probe structural degradation, nodes are removed in steps of $1\%$. After each step, we retain the largest connected component (LCC), recompute $\beta_{\Delta}$, and evaluate the new effective steady state.
For the Erd\H{o}s--R\'enyi network, we use $N=500$ and $p=0.1$, so that $5$ nodes are removed at each step. The comparison in Fig. \ref{fig:validation_gene} (a) shows that the reduced model follows the full-network branches very closely for both $0.01$ and $10.0$ initial conditions. The transition from the low-expression branch to the high-expression branch is captured well, and the steady-state growth with $\beta_{\Delta}$ is reproduced over the full range shown.
For the Barab\'asi--Albert network, we again take $N=500$, with $m=12$, where $m$ is the number of edges attached by each new node during growth. Fig. \ref{fig:validation_gene} (b) shows that the reduced model still reproduces the main branch structure and the overall increase of the steady state with $\beta_{\Delta}$. The agreement remains good, although the deviation is slightly larger than in the ER case.
Fig. \ref{fig:validation_gene} (c) shows the bifurcation diagram of the reduced one-dimensional system. The stable and unstable branches are clearly separated, and the transition from the low-expression state to the high-expression state is preserved in the reduced dynamics.
For the real-network test, we use the \textit{C. elegans} integrated functional interaction network from the \href{https://interactome.dfci.harvard.edu/C_elegans/index.php?page=download}{CCSB Interactome Database}. The dataset combines WI8, literature, microarray, phenotype, interolog, and genetic evidence, and contains $6125$ nodes. Here, about $61$ nodes are removed at each step. As shown in Fig. \ref{fig:validation_gene} (d), the reduced model captures the main steady-state trend and the branch transition of the real network. The agreement is strong at the level of the effective branch structure.

\begin{figure}[H]
    \centering
    \includegraphics[width=0.6\linewidth]{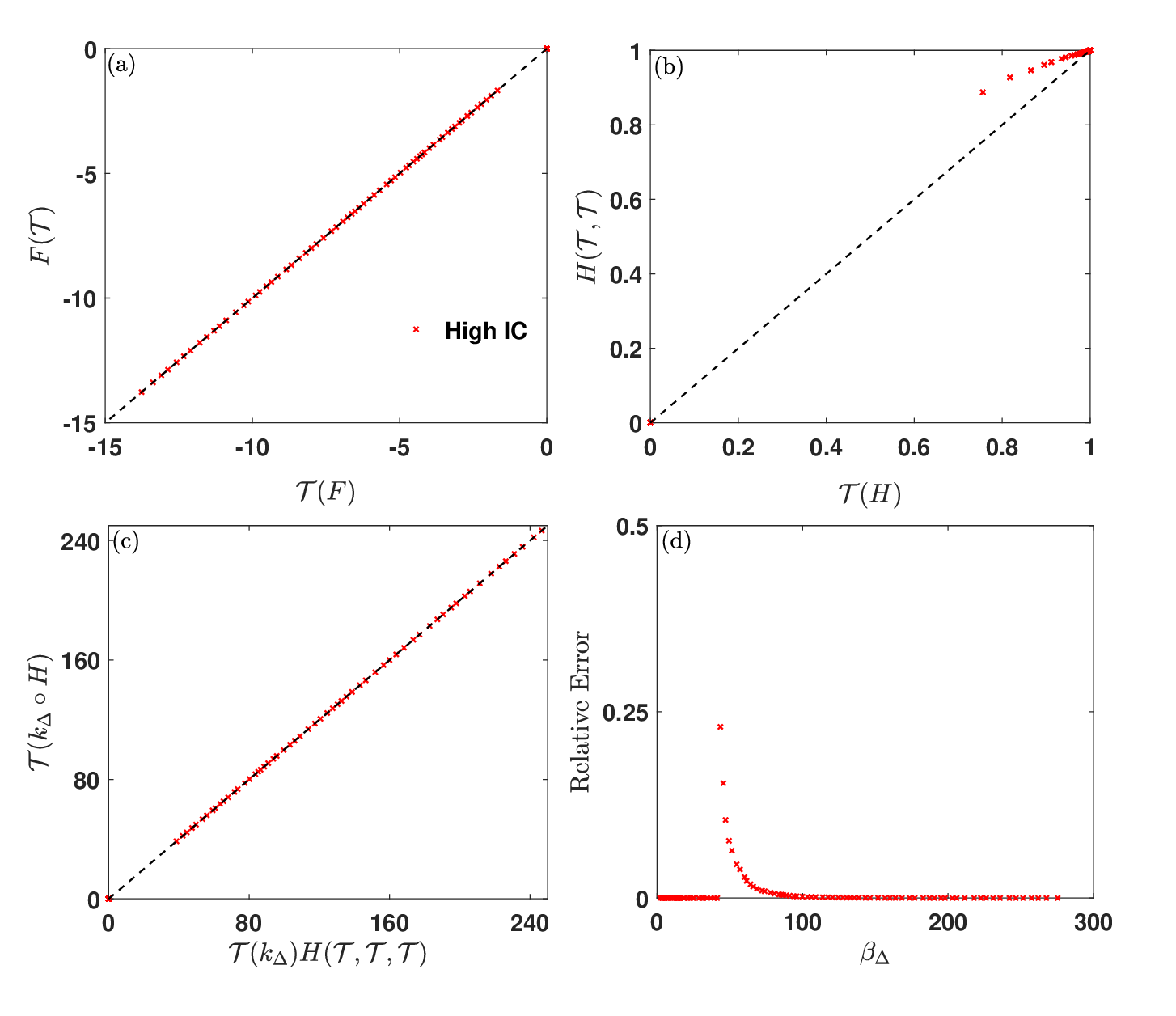}
  % Figure 9
\caption{\textbf{Approximation analysis for the ER gene regulatory network.} Panels (a)–(c) compare the full-network terms entering the approximations with their reduced counterparts. The points lie close to the ideal diagonal in all three panels, with only a small deviation in panel (b) at larger values. Panel (d) shows the relative error of the effective steady state as a function of $\beta_{\Delta}$. The error is negligible over most of the range, apart from a narrow peak in the high-initial-condition branch near the transition.}
       \label{GENE_ER}
\end{figure}

\begin{figure}[H]
    \centering
    \includegraphics[width=.6\linewidth]{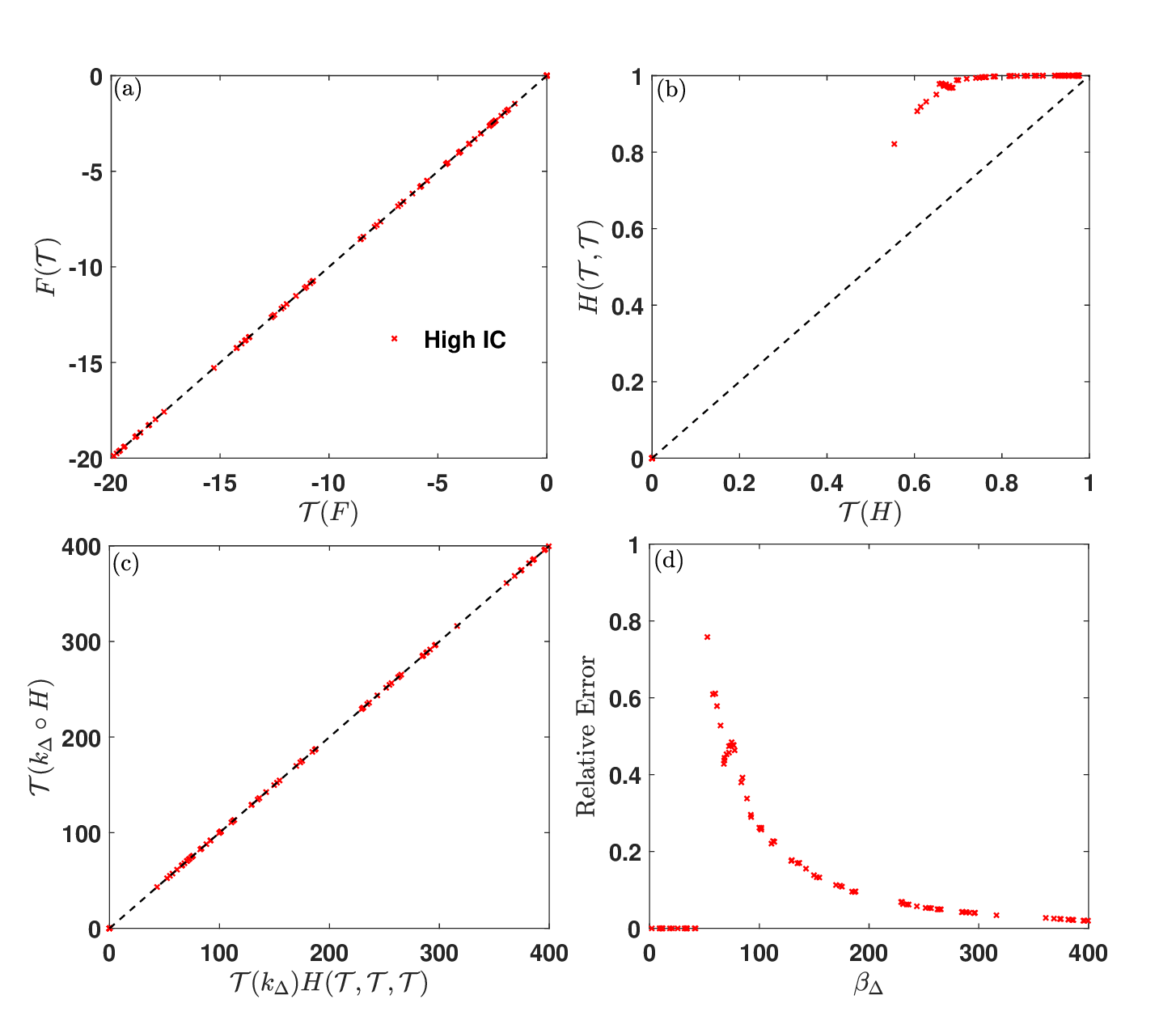}
   % Figure 10
\caption{\textbf{Approximation analysis for the BA gene regulatory network.} Panels (a)–(c) compare the full-network terms entering the approximations with their reduced counterparts. Panels (a) and (c) remain close to the ideal diagonal, whereas panel (b) shows a more visible departure from the ideal relation over part of the range. Panel (d) shows the relative error of the effective steady state as a function of $\beta_{\Delta}$. The error is small over most of the range, with a narrow peak in the high-initial-condition branch near the transition.}
\label{gene_BA}
\end{figure}

\begin{figure}[H]
    \centering
    \includegraphics[width=.6\linewidth]{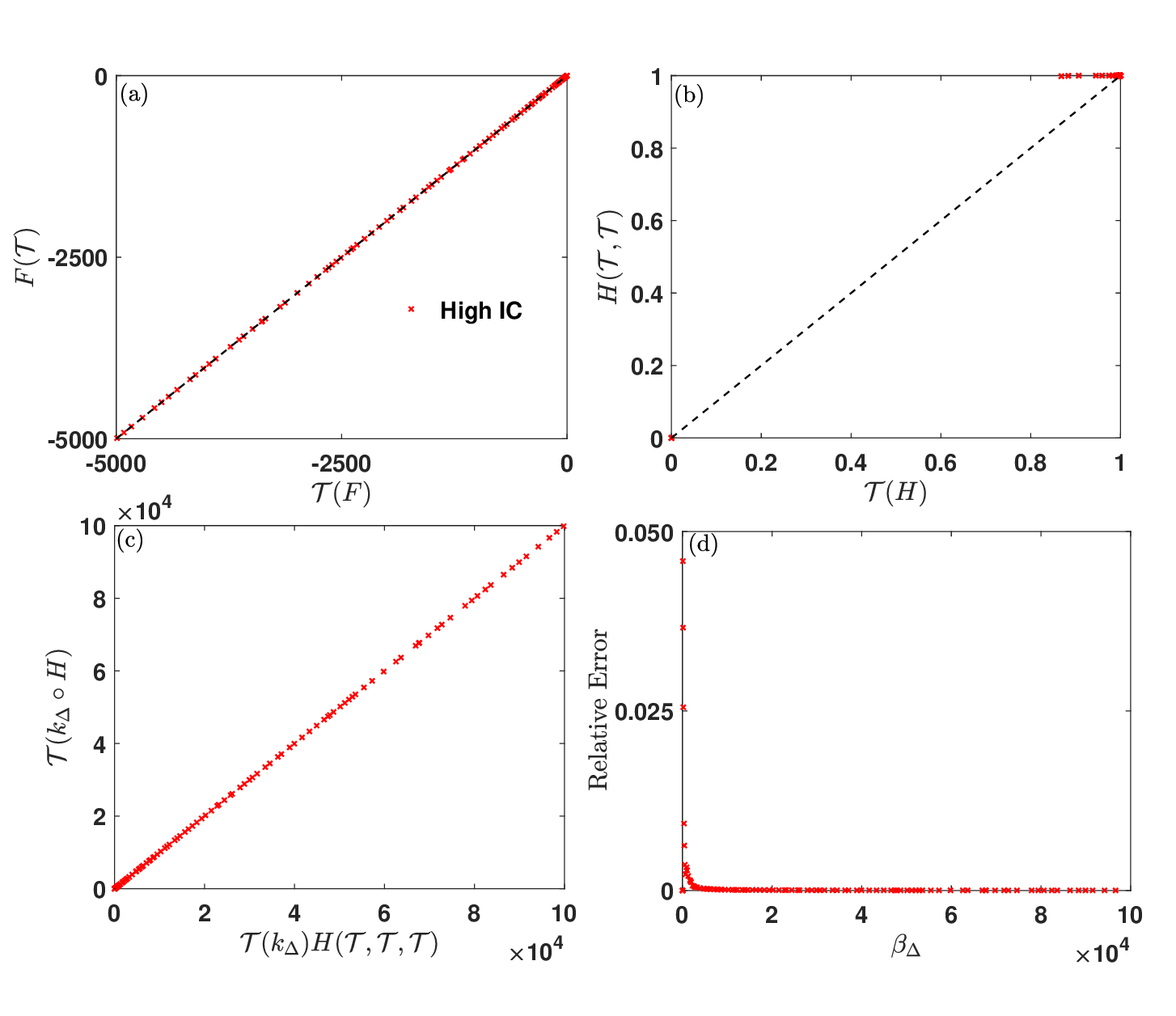}
    
    \caption{\textbf{Approximation analysis for the real gene regulatory network.} Panels (a)–(c) compare the full-network quantities entering the approximations with their reduced counterparts. In panels (a) and (c), most points remain close to the ideal diagonal, while panel (b) shows a larger deviation at higher values. Panel (d) shows the relative error of the effective steady state as a function of $\beta_{\Delta}$. The error is largest at small $\beta_{\Delta}$ and decreases rapidly, remaining small over most of the range. Overall, the reduction captures the main effective behaviour of the real network, although the agreement is weaker than in the synthetic cases.}
\label{GENE_REAL}
\end{figure}

The approximation-level behavior is shown in the Approximation Figs. \ref{GENE_ER} - \ref{GENE_REAL}. For the ER network Fig. \ref{GENE_ER}, the closures remain close to the ideal relations over almost the full range, and the relative error is negligible except for a very narrow peak. For the BA network Fig. \ref{gene_BA}, the same overall pattern holds, although the interaction closure shows a larger deviation from the ideal line. For the real network Fig. \ref{GENE_REAL}, the approximation quality is again mixed: the intrinsic and factorization terms remain close to the ideal relation, while the interaction closure shows a visible spread, especially at larger values. Even so, the relative error remains small, which is consistent with the good agreement seen in Fig. \ref{fig:validation_gene} (d).

\subsection{SIS model}

We finally test the reduction for the susceptible--infected--susceptible (SIS) model with purely higher-order interactions:
\begin{equation}
\dot{x}
=
-\mu x
+
\gamma\beta_{\Delta}(1-x)x^2.
\end{equation}
Here, $x$ denotes the effective infected fraction. The first term represents recovery, with rate $\mu$, and the second term represents higher-order infection, with rate $\gamma$. The parameter $\beta_{\Delta}$ is the effective higher-order interaction strength associated with the network. Throughout this subsection, we use
\[
\mu=1,\qquad \gamma=0.05.
\]

The steady states satisfy
\begin{equation}
x\left[-\mu+\gamma\beta_{\Delta}x(1-x)\right]=0.
\end{equation}
Thus, $x=0$ is always a fixed point. For $x\neq 0$, the nonzero steady states satisfy
\begin{equation}
\gamma\beta_{\Delta}x^2-\gamma\beta_{\Delta}x+\mu=0,
\end{equation}
which gives
\begin{equation}
x=\frac{1}{2}\left(1\pm\sqrt{1-\frac{4\mu}{\gamma\beta_{\Delta}}}\right).
\end{equation}
These nonzero fixed points exist only when
\begin{equation}
\gamma\beta_{\Delta}\geq 4\mu.
\end{equation}
Hence, the reduced SIS system always has the zero fixed point, while the nonzero branch appears only above the threshold $\gamma\beta_{\Delta}=4\mu$  \cite{ghosh2023chaos}.

For the full network dynamics, we integrate from low and high initial conditions ($0.01$, $10$) until a steady state is reached. After each node-removal step, we retain the largest connected component (LCC), recompute $\beta_{\Delta}$, and evaluate the corresponding effective steady state.

\begin{figure}[H]
    \centering
    \includegraphics[width=1\linewidth]{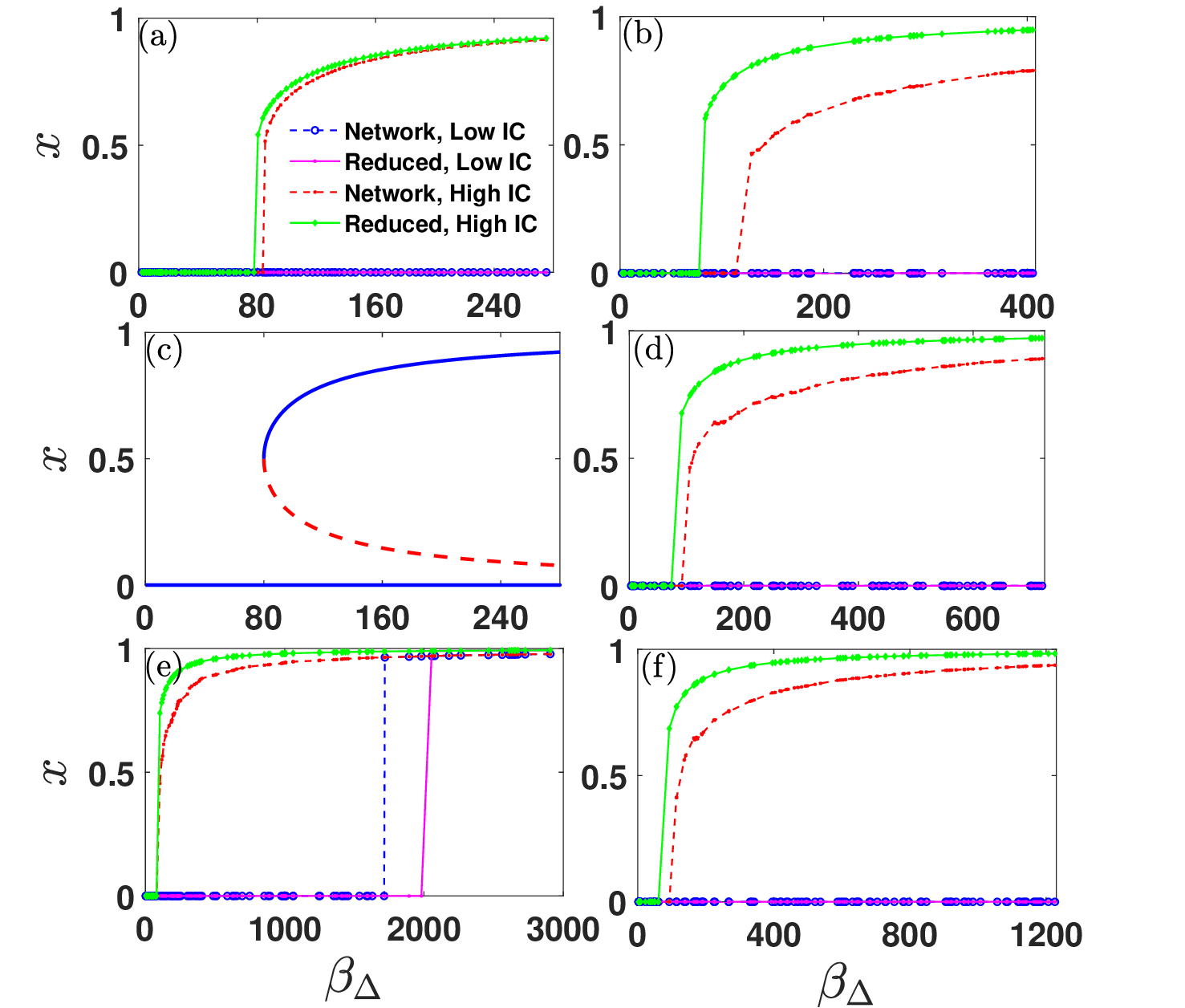}
 \caption{\textbf{SIS model with higher-order interactions.} Steady-state value $x$ as a function of the effective higher-order interaction strength $\beta_{\Delta}$. Panel (a) compares the full-network simulations with the one-dimensional reduction for an Erd\H{o}s--R\'enyi network with $p=0.1$. Panels (b), (d), and (f) show the corresponding results for Barab\'asi--Albert networks with $m=12$, $m=16$, and $m=20$, respectively, where $m$ is the number of edges attached by each new node during growth. Panel (c) shows the bifurcation diagram of the reduced system, including the stable and unstable branches. Panel (e) presents the result for the email-Eu-core network, constructed from the undirected projection of the original directed dataset. The reduction captures the main transition from the zero state to the nonzero state and reproduces the branch structure with improving agreement as $m$ increases.}
\label{fig:validation_sis}
\end{figure}

Fig.~\ref{fig:validation_sis} (a) shows the result for an Erd\H{o}s--R\'enyi (ER) network with $p=0.1$. The reduced model follows the full-network branches well and captures the transition from the zero state to a nonzero infected state as $\beta_{\Delta}$ increases.

For Barab\'asi-Albert (BA) networks, we consider three values of the growth parameter: $m=12$, $m=16$, and $m=20$, where $m$ is the number of edges attached by each new node during growth. The corresponding results are shown in Figs.~\ref{fig:validation_sis} (b),(d) and  (f). In all three cases, the reduced model reproduces the main branch structure. The agreement becomes stronger as $m$ increases, indicating that the reduction performs better in the more connected BA networks.

Fig.~\ref{fig:validation_sis} (c) shows the bifurcation diagram of the reduced one-dimensional SIS system. The zero fixed point, the stable nonzero branch, and the unstable intermediate branch are clearly identified, showing that the reduced dynamics preserve the threshold and bistable structure of the SIS model.

For the real-network test, we use the email communication network from the \href{https://snap.stanford.edu/data/email-Eu-core.html}{SNAP email-Eu-core dataset}, which contains $1005$ nodes. Since the original network is directed, we first convert it to its undirected projection and then construct the higher-order interaction structure from that network. Fig.~\ref{fig:validation_sis} (e) shows that the reduced model captures the main transition from the zero state to the nonzero state and follows the effective steady-state trend reasonably well.

To further examine the threshold behavior, Fig.~\ref{fig:placeholder_sis} shows the steady state as a function of the infection rate $\gamma$ for the ER network. The reduced model reproduces the bistable region and the onset of the nonzero branch, consistent with the threshold structure predicted by the reduced equation.

The approximation-level behavior is shown in Appendix Figs.~\ref{SIS_ER} -~\ref{SIS_BA_12}. For the ER network Fig.~\ref{SIS_ER}, the closures remain close to the ideal relations, and the relative error is negligible except for a very narrow peak near the transition. For the BA networks Figs.~\ref{SIS_BA_20}--~\ref{SIS_BA_12}, the same overall pattern holds; the interaction closure shows a modest spread, and the error becomes smaller as $m$ increases. For the real network Fig.~\ref{SIS_REAL}, the approximation quality is mixed, with a broader deviation in the interaction closure and a larger peak in the low-IC error. Even so, the reduced model still captures the main branch structure and threshold behavior seen in the full system.

\section{Dependence on system parameter}
\begin{figure}[H]
    \centering
    \includegraphics[width=\linewidth]{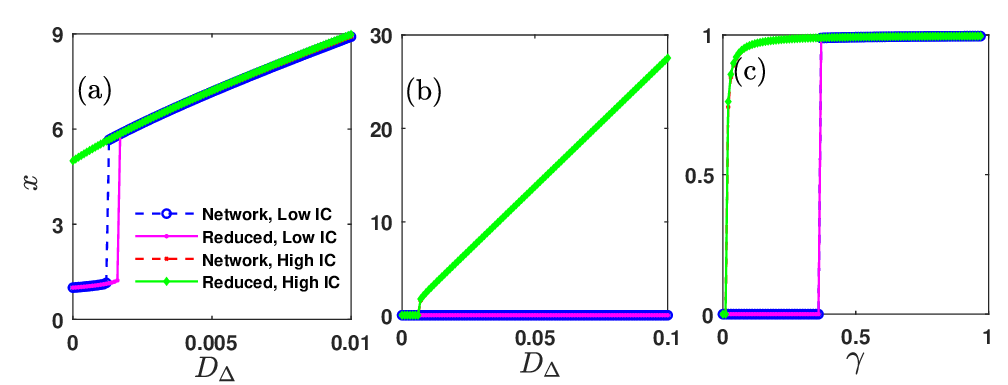}
  \caption{System-parameter dependence of the network and reduced dynamics on an Erdős-Rényi network of 500 nodes. (a) For the double-well system, varying $D_{\Delta}$ from 0 to 0.01 shows good agreement between the network and reduced descriptions. (b) For the gene system, varying $D_{\Delta}$ from 0 to 0.1  (c). For the $\gamma$-dependent system, varying $\gamma$ from 0 to 1 again gives good overall agreement.}
    \label{fig:placeholder_sis}
\end{figure}

To examine how the accuracy of the reduced description depends on the choice of the control parameter, we varied it across three representative dynamical models while keeping the underlying network fixed. This helps us test whether the one-dimensional reduction remains reliable across different choices of control parameters in the chosen dynamical system. We find that the reduced model captures the double-well, gene dynamics, and the SIS model. These results indicate that the reduction is independent of the choice of system parameters. In Fig. \ref{fig:placeholder_sis} (a), (b), and (c), we respectively show the variation of control parameter $D_{\Delta}$ in the double well system, $D_{\Delta}$ in the gene regulatory system, and $
\gamma$ in the SIS model.

\section*{5. Discussion}

In this work, we developed a one-dimensional effective-state reduction framework for dynamical systems on networks with purely higher-order interactions. By introducing the effective higher-order interaction strength parameter $\beta_{\Delta}$, the framework extends the effective-state idea to systems driven by triangular interaction structures. This reduction provides a simple, low-dimensional description of the collective dynamics while retaining the main role of higher-order network organization.

Through analytical derivations, we identified the main approximations underlying the reduction and expressed their validity through the ratios $R_2$, $R_3$, and $R_4$. The analysis shows that the reduction accuracy is influenced primarily by the homogeneity of node states, the quality of the higher-order mean-field closure, and, in some cases, the dependence of node states on higher-order degree. For the double-well and gene-regulatory systems, the results indicate that state homogeneity is the main requirement for accurate reduction. In contrast, in the SIS system, interaction closure and degree dependence play a more prominent role. Overall, these results suggest that the success of the reduction is determined more strongly by the distribution of node states than by the network structure alone.\\
We tested the framework on three dynamical systems: the double-well potential system, a nonlinear gene-regulatory system, and the SIS epidemic model with purely higher-order interactions. To support the interpretation of these numerical results, we also report the structural properties of all synthetic and real networks used in this study in the Appendix: see Tables~\ref{tab:structural_basic} and~\ref{tab:structural_hoi}, including degree heterogeneity, triangular-degree statistics, assortativity, modularity, and clustering. Across Erd\H{o}s-R\'enyi and Barab\'asi-Albert networks, the reduced model reproduces the main steady-state branches and transition structures with good agreement. For the real-world networks, the agreement is more mixed, reflecting stronger structural heterogeneity and community organization, but the reduced description still captures the main effective trend and the principal branch transition. These results support the usefulness of the proposed framework as a low-dimensional tool for studying effective steady states and resilience-related transitions in higher-order networked systems.\\
The framework also highlights the role of intrinsic dynamics in shaping reduction accuracy. In the SIS model, the approximation associated with the linear recovery term is satisfied exactly, whereas in the double-well and gene-regulatory systems, the nonlinear intrinsic terms make the reduction more sensitive to state heterogeneity. This contrast shows that the quality of a one-dimensional description depends not only on the higher-order interaction structure but also on the form of the underlying node dynamics.\\
The role of state heterogeneity can also be quantified within the present framework. In networks with degree variation, community structure, or uneven participation in higher-order motifs, different nodes may receive different local higher-order inputs. The approximation ratios \(R_2\), \(R_3\), and \(R_4\), together with the relative error between the effective state calculated from the states of the original network and the reduced steady state, provide practical diagnostics for this effect. When these quantities remain close to their ideal values, the one-dimensional reduction gives an accurate effective description. This also suggests a natural extension of the framework: for strongly structured networks, one may use a small number of effective states based on degree classes, communities, or higher-order motif participation.\\
 The reduction is most accurate when node states remain sufficiently homogeneous and when a single effective parameter can represent the higher-order input experienced by different nodes. Its accuracy may decrease in degree-degree correlated networks, strongly heterogeneous networks, or in dynamics that amplify deviations between individual node states and the effective state. Although the present framework is formulated for triangular interactions through the effective parameter ($\beta_{\Delta}$), the same reduction idea can, in principle, be extended to four-node motifs or general ($m$)-body hyperedges. In such cases, ($\beta_{\Delta}$) would need to be replaced or supplemented by effective parameters for the relevant higher-order motifs, and the closure conditions would need to be generalized to the corresponding ($m$)-body interaction functions. This extension to general hypergraph or simplicial-
complex dynamics is a natural direction for future work.

\vspace*{0.15pt}

\enlargethispage{20pt}

\noindent\textbf{Declaration of AI use.} 
During the preparation of this work, the authors used language refinement tools to refine the English style of the presentation alongside grammatical corrections. After using this tool/service, the authors reviewed and edited the content as needed and take full responsibility for the publication's content.\\
\textbf{Acknowledgment.} CH acknowledges support from ARNF India \\(Grant Number ANRF/ECRG/2024/000207/PMS). AT and PK acknowledge Mukesh Tiwari for valuable and fruitful discussions.

\section{Appendix}
The appendix provides a detailed approximation-level assessment of the one-dimensional reduction for the three dynamical systems considered in the main text. For each system and network type, we compare the full-network quantities entering the approximations $(A_2)$--$(A_4)$ with their reduced counterparts and report the corresponding relative error of the effective steady state. These additional results complement the main bifurcation plots by showing where the reduction remains close to the ideal closure relations and where visible deviations appear.
\subsection{Double Well}

\begin{figure}[H]
    \centering
    \includegraphics[width=.6\linewidth]{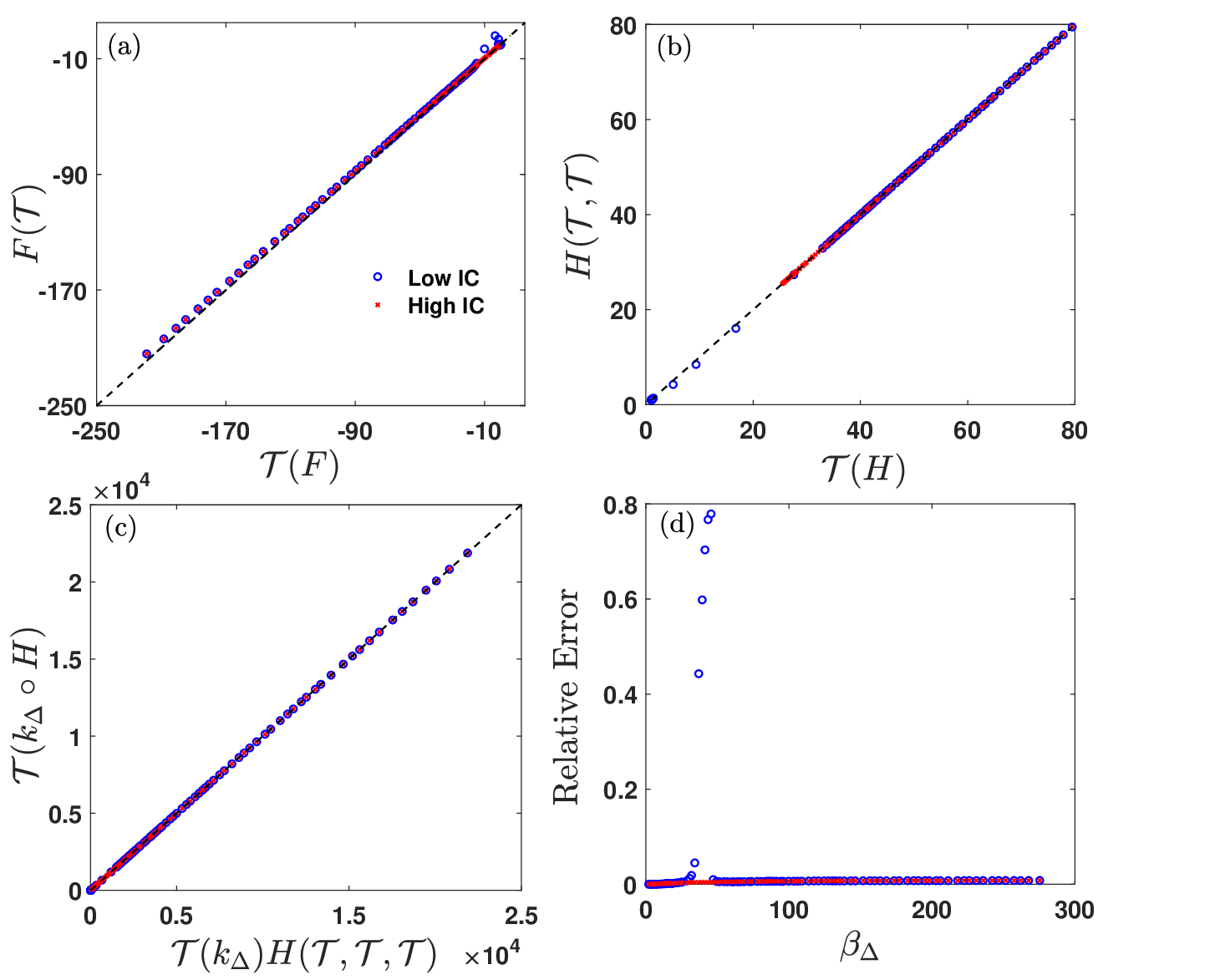}
 \caption{\textbf{Approximation analysis for the ER double-well network.} Panels (a)–(c) compare the full-network terms entering the approximations with their reduced counterparts. Most points lie close to the ideal diagonal, indicating that the approximations remain accurate over most of the parameter range. Panel (d) shows the relative error of the effective steady state as a function of $\beta_{\Delta}$. The error is small except for a narrow peak near the transition in the low-initial-condition branch.}

   \label{DW_ER}
\end{figure}
\begin{figure}[H]
    \centering
    \includegraphics[width=.6\linewidth]{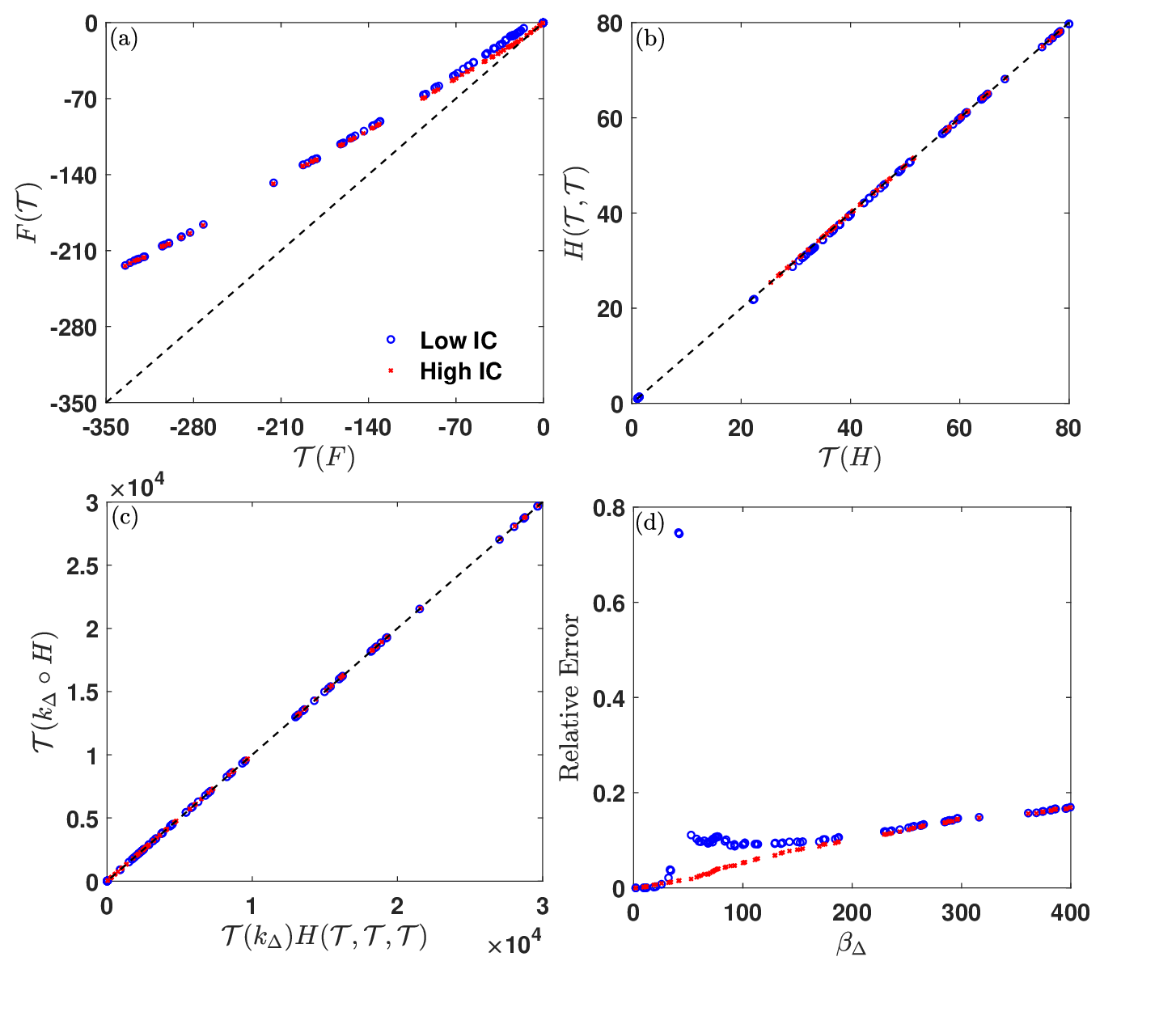}
    % Figure 7
\caption{\textbf{Approximation analysis for the BA double-well network.} Panels (a)–(c) compare the full-network terms entering the approximations with their reduced counterparts. Panels (b) and (c) remain close to the ideal diagonal, whereas panel (a) shows a larger spread, especially at smaller values. Panel (d) shows the relative error of the effective steady state as a function of $\beta_{\Delta}$. The error is larger than in the ER case, particularly for the low-initial-condition branch, but the reduction still captures the main effective behaviour of the system.}

   \label{DW_BA}
\end{figure}

\begin{figure}[H]
    \centering
    \includegraphics[width=0.6\linewidth]{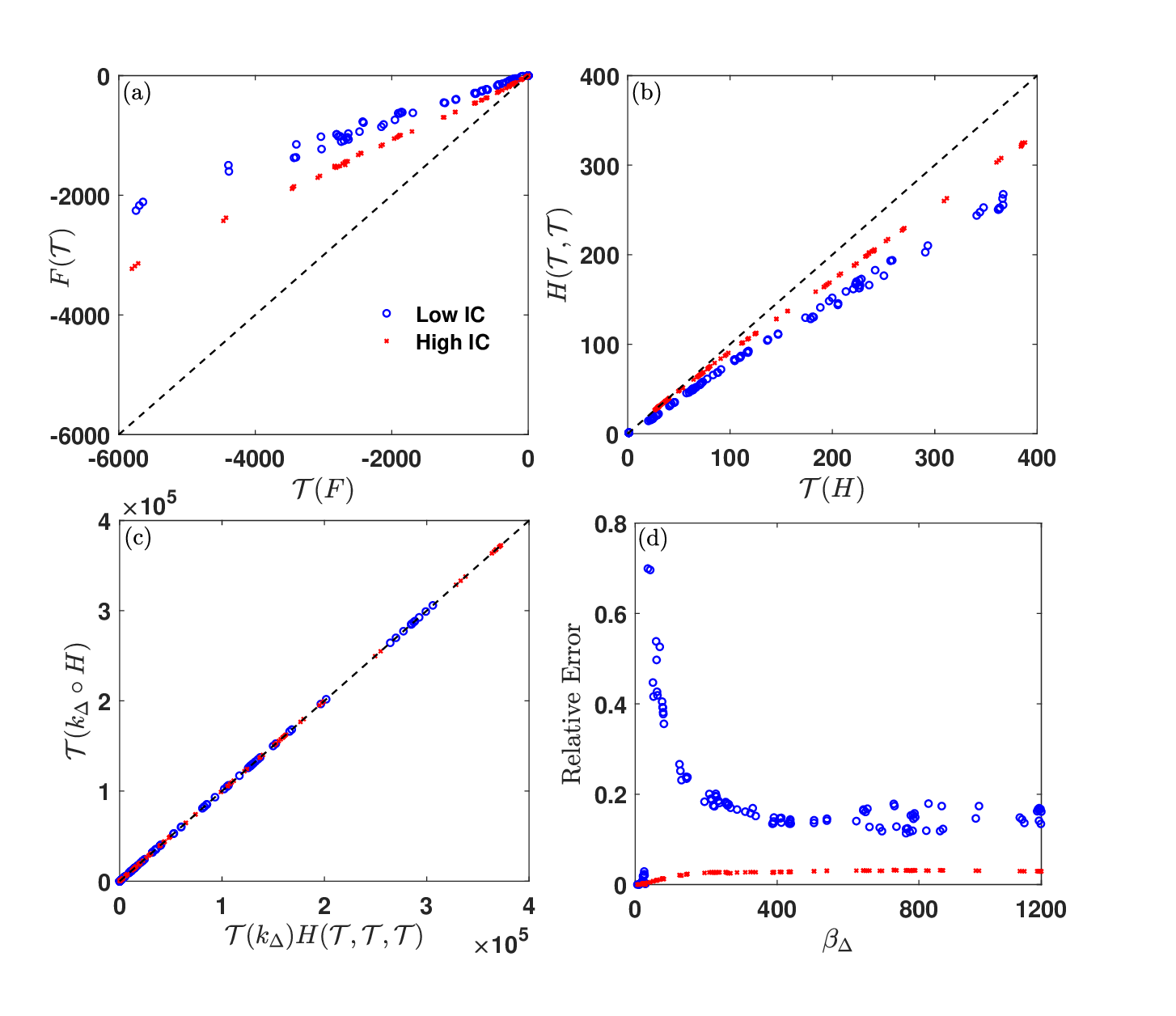}
   % Figure 8
\caption{\textbf{Approximation analysis for the real double-well network.} Panels (a)–(c) compare the full-network terms entering the approximations with their reduced counterparts. Panel (c) remains close to the ideal diagonal, whereas panels (a) and (b) show visibly larger deviations, especially at larger values. Panel (d) shows the relative error of the effective steady state as a function of $\beta_{\Delta}$. The high-initial-condition branch remains relatively close to the reduced description, while the low-initial-condition branch shows a larger error over a broader range.}
   \label{DW_REAL}
\end{figure}
\subsection{Gene Regulatory System}

\subsection{SIS}
\begin{figure}[H]
    \centering
    \includegraphics[width=.6\linewidth]{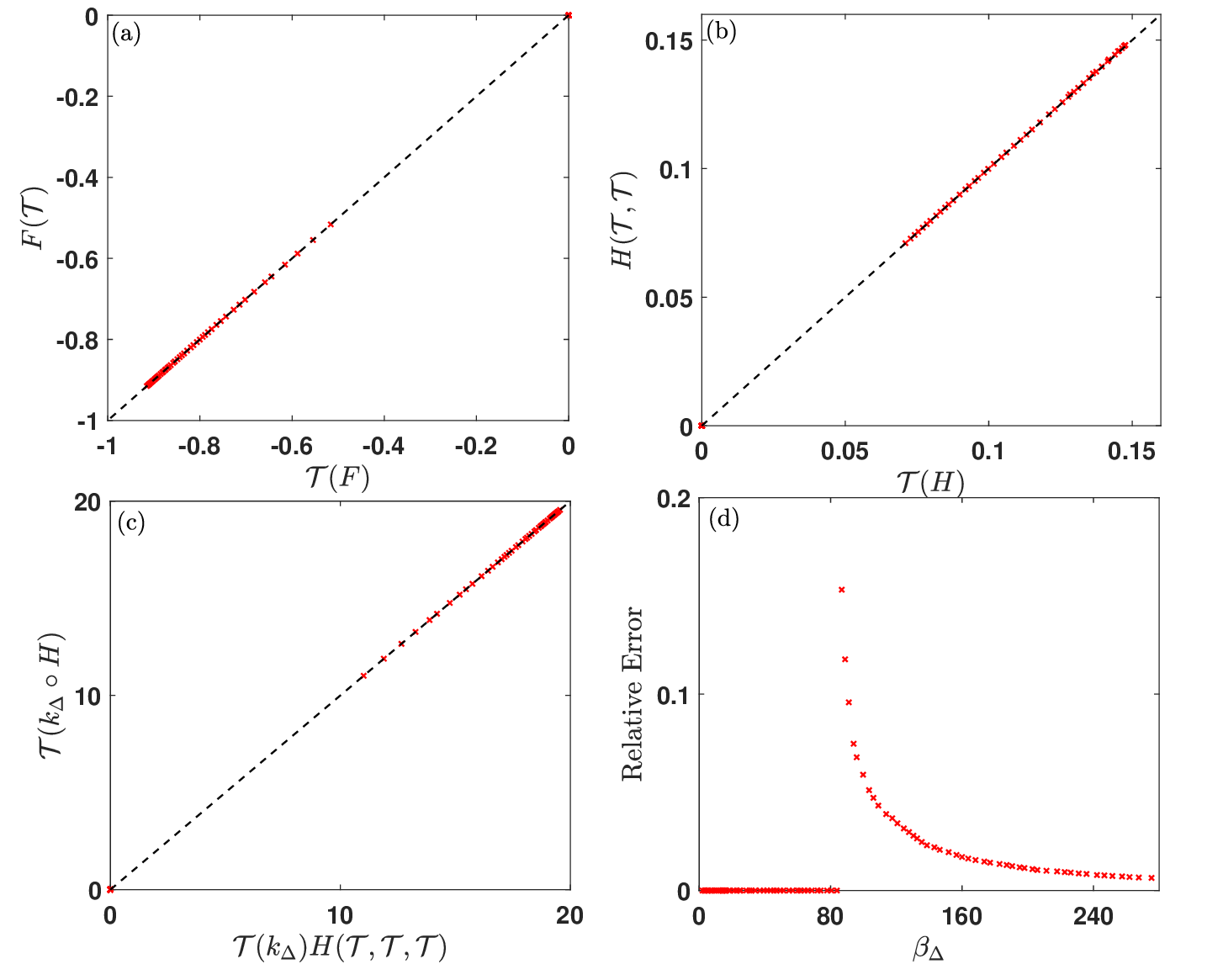}
    \caption{\textbf{Approximation analysis for the ER SIS network.} Panels (a)–(c) compare the full-network quantities entering the approximations with their reduced counterparts. In all three panels, the points lie close to the ideal diagonal, indicating that the approximations remain accurate over most of the parameter range. Panel (d) shows the relative error of the effective steady state as a function of $\beta_{\Delta}$. The error is negligible except for a narrow peak near the transition.}
       \label{SIS_ER}
\end{figure}
\begin{figure}[H]
    \centering
    \includegraphics[width=0.6\linewidth]{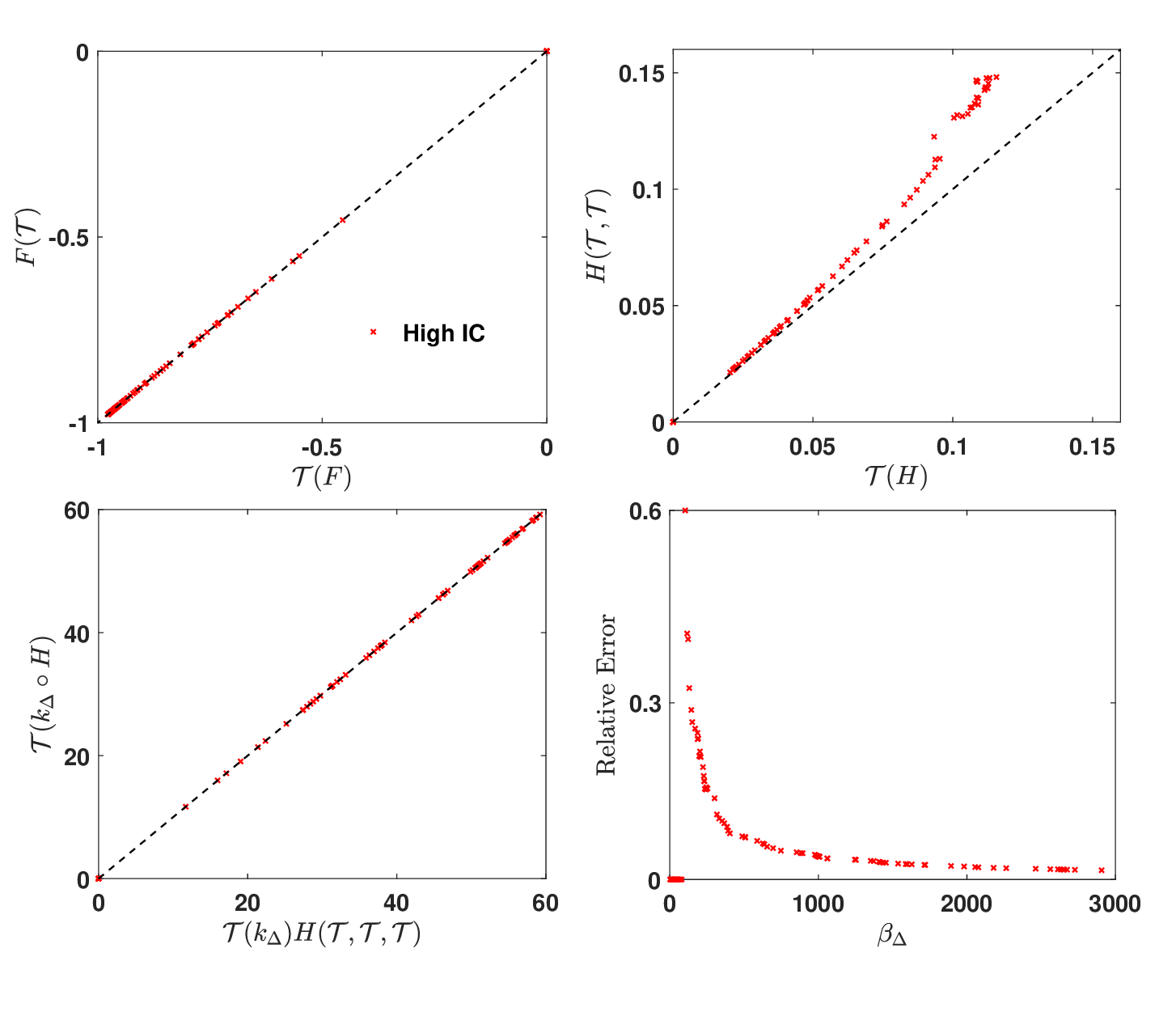}
  \caption{\textbf{Approximation analysis for the real SIS network.} Panels (a)–(c) compare the full-network quantities entering the approximations with their reduced counterparts. Panels (a) and (c) remain close to the ideal diagonal, whereas panel (b) shows a more visible deviation at larger values. Panel (d) shows the relative error of the effective steady state as a function of $\beta_{\Delta}$. The error is larger near the transition and then decreases, indicating that the reduction captures the main effective behaviour but with weaker agreement than in the ER case.}
    \label{SIS_REAL}
\end{figure}
\begin{figure}[H]
    \centering
    \includegraphics[width=0.6\linewidth]{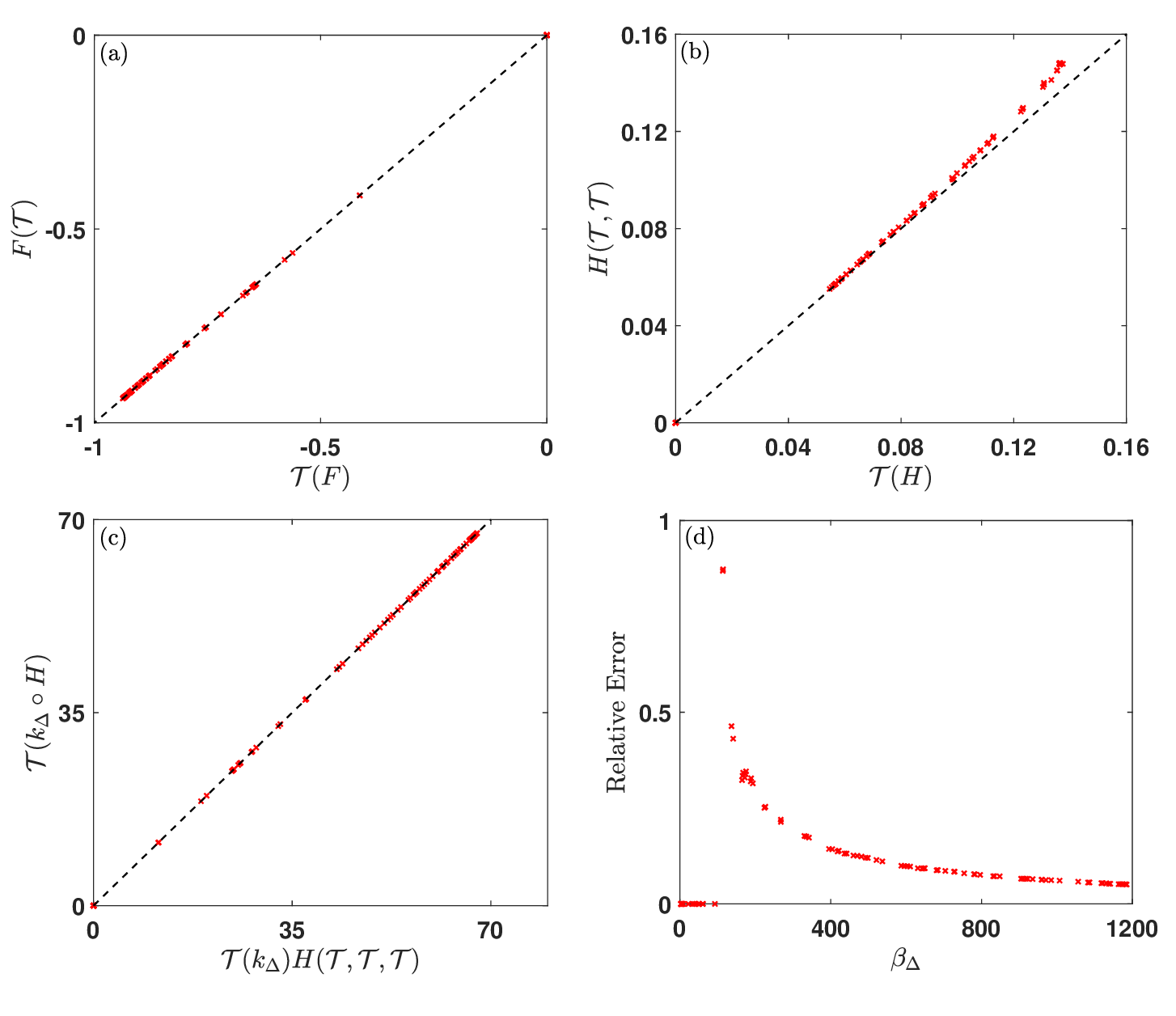}
\caption{\textbf{Approximation analysis for the BA SIS network with $m=20$.} Panels (a)–(c) compare the full-network quantities entering the approximations with their reduced counterparts. Most points lie close to the ideal diagonal, indicating that the approximations remain accurate over most of the parameter range, with only a small spread in panel (b). Panel (d) shows the relative error of the effective steady state as a function of $\beta_{\Delta}$. The error is negligible except for a narrow peak near the transition, consistent with the good agreement seen in the main bifurcation plot.}
    \label{SIS_BA_20}
\end{figure}
\begin{figure}[H]
    \centering
    \includegraphics[width=0.6\linewidth]{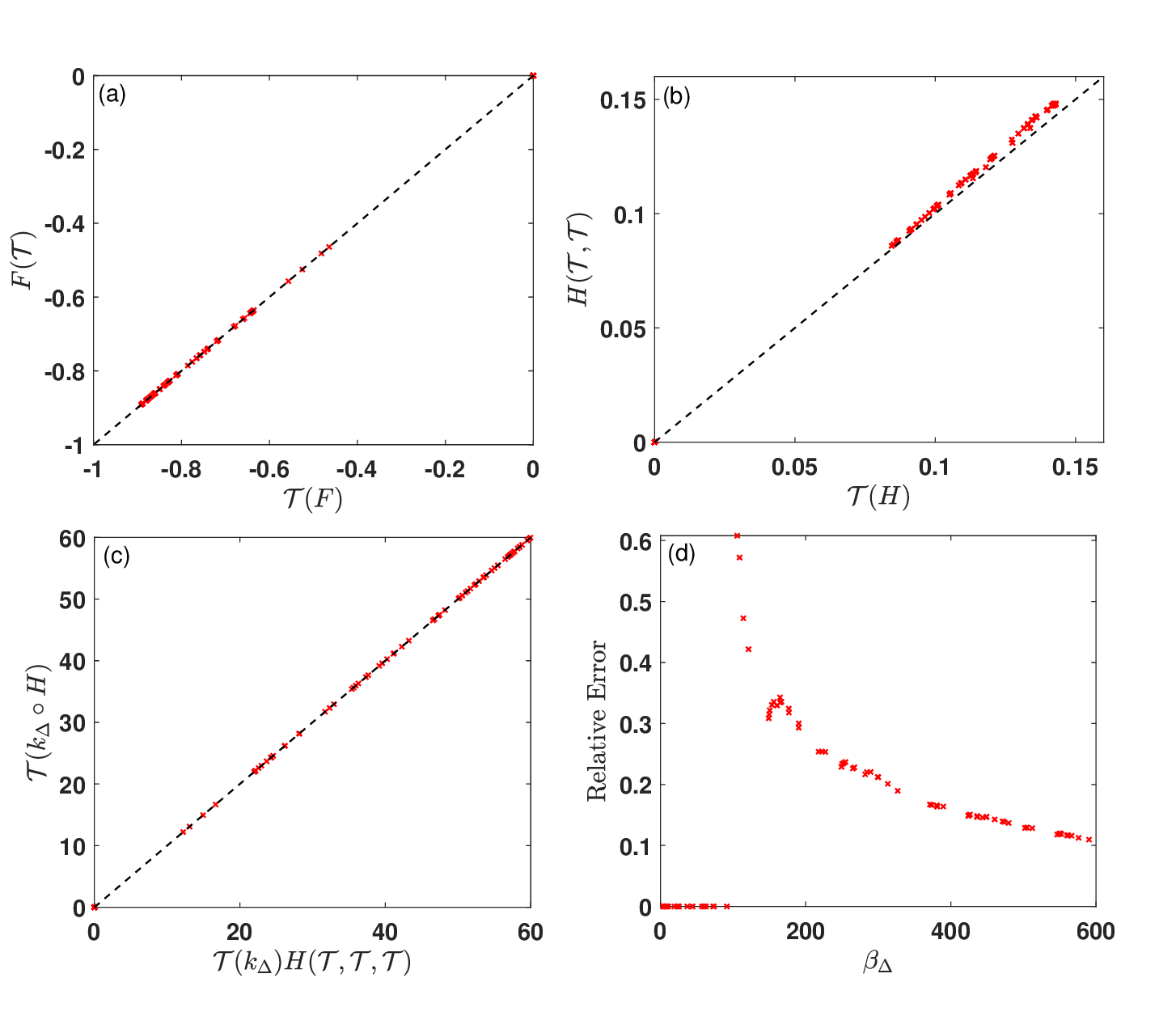}
\caption{\textbf{Approximation analysis for the BA SIS network with $m=16$.} Panels (a)–(c) compare the full-network quantities entering the approximations with their reduced counterparts. The points remain close to the ideal diagonal in all three panels, with only a small deviation in panel (b) at larger values. Panel (d) shows the relative error of the effective steady state as a function of $\beta_{\Delta}$. The error is negligible except for a narrow peak near the transition, indicating good agreement between the reduced and full-network results over most of the parameter range.}
\label{SIS_BA_16}
\end{figure}
\begin{figure}[H]
    \centering
    \includegraphics[width=0.6\linewidth]{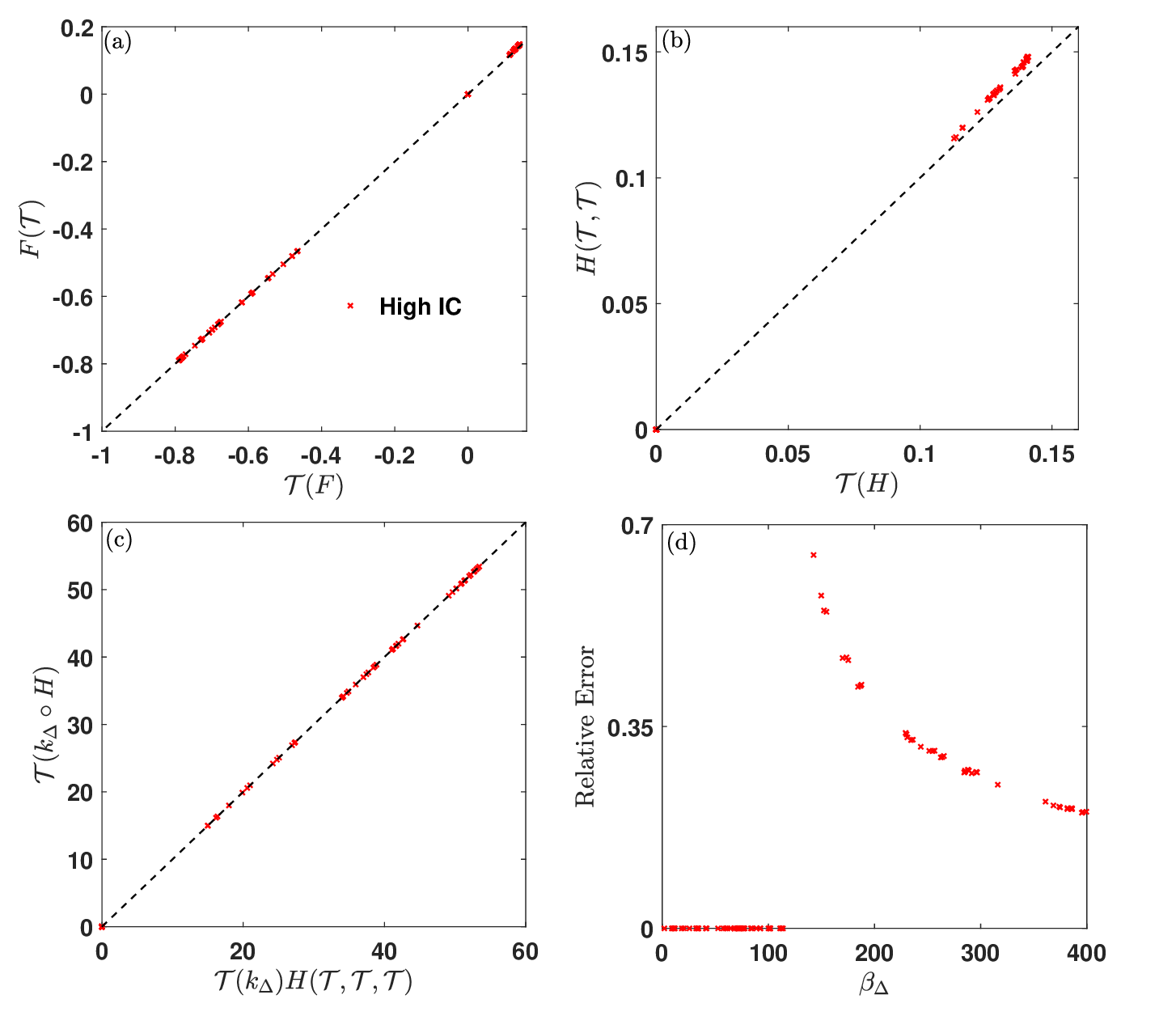}
    \caption{\textbf{Approximation analysis for the BA SIS network with $m=12$.} Panels (a)–(c) compare the full-network quantities entering the approximations with their reduced counterparts. Most points remain close to the ideal diagonal, although panel (b) shows a slightly larger spread than in the cases with higher $m$. Panel (d) shows the relative error of the effective steady state as a function of $\beta_{\Delta}$. The error is concentrated near the transition region and is otherwise small, showing that the reduction captures the main effective behaviour of the system.}
    \label{SIS_BA_12}
\end{figure}
\subsection{Structural Characterization of Synthetic and Real Networks}

To characterize the synthetic and real networks used in the numerical validation, we report several structural diagnostics. Here, $N$ denotes the number of nodes, $E$ denotes the number of pairwise edges, and Tri. denotes the number of triangular interactions present in the network. The ordinary degree of node $i$ is denoted by $k_i$, with $\langle k\rangle$ and $\langle k^2\rangle$ representing the first and second moments of the degree distribution, respectively. The quantity $CV(k)$ denotes the coefficient of variation of the ordinary degree distribution and measures degree heterogeneity. For higher-order interactions, $k_i^{\Delta}$ denotes the triangular degree of node $i$, defined as $k_i^{\Delta}=\sum_{j,l}T_{ijl}$, where $T_{ijl}$ represents a triangular interaction among nodes $i$, $j$, and $l$. Accordingly, $\langle k_{\Delta}\rangle$ and $CV(k_{\Delta})$ describe the mean and heterogeneity of the triangular-degree distribution. The effective higher-order structural parameter is given by $\beta_{\Delta}=\langle k_{\Delta}^{2}\rangle/\langle k_{\Delta}\rangle$. The quantity $r$ denotes ordinary pairwise degree assortativity, while $r_{\Delta}$ denotes higher-order assortativity computed from the triangular-degree distribution \cite{landry2022hypergraph}. Finally, $Q$ denotes the modularity obtained using the Louvain community-detection algorithm, Comm. is the corresponding number of detected communities, and Avg. Clust. denotes the average clustering coefficient of the network.

\newpage
\begin{table}[H]
\centering
\small
\setlength{\tabcolsep}{2pt}
\renewcommand{\arraystretch}{1.10}
\begin{tabular*}{\textwidth}{@{\extracolsep{\fill}}|l|l|c|c|c|c|c|c|c|@{}}
\toprule
Model & Network & $N$ & $E$ & Tri. & $\langle k\rangle$ & $\langle k^2\rangle$ & $CV(k)$ & Avg. Clust. \\
\midrule
SIS & ER & 500 & 12474 & 20835 & 49.8960 & 2534.0400 & 0.1336 & 0.1008 \\
SIS & BA12 & 500 & 5856 & 7098 & 23.4240 & 851.7000 & 0.7431 & 0.1092 \\
SIS & BA16 & 500 & 7744 & 14273 & 30.9760 & 1400.2360 & 0.6777 & 0.1289 \\
SIS & BA20 & 500 & 9600 & 25423 & 38.4000 & 2105.8320 & 0.6543 & 0.1559 \\
SIS & Email & 986 & 16064 & 105461 & 32.5842 & 2432.6166 & 1.1363 & 0.4071 \\
\midrule
DW & ER & 500 & 12394 & 20373 & 49.5760 & 2504.3680 & 0.1377 & 0.0995 \\
DW & BA12 & 500 & 5856 & 7023 & 23.4240 & 857.1040 & 0.7497 & 0.1105 \\
DW & Collaboration & 4158 & 13422 & 47779 & 6.4560 & 116.0851 & 1.3361 & 0.5569 \\
\midrule
Gene & ER & 500 & 12558 & 21053 & 50.2320 & 2568.9760 & 0.1346 & 0.1001 \\
Gene & BA12 & 500 & 5856 & 7198 & 23.4240 & 866.6960 & 0.7613 & 0.1139 \\
Gene & Yeast & 4441 & 12864 & 3750 & 5.7933 & 517.9144 & 3.7989 & 0.0834 \\
\bottomrule
\end{tabular*}
\caption{Basic structural properties of the synthetic and real networks considered for each dynamical model.}
\label{tab:structural_basic}
\end{table}

\begin{table}[t]
\centering
\small
\setlength{\tabcolsep}{2pt}
\renewcommand{\arraystretch}{1.10}
\begin{tabular*}{\textwidth}{@{\extracolsep{\fill}}|l|l|c|c|c|c|c|c|c|@{}}
\toprule
Model & Network & $\langle k_{\Delta}\rangle$ & $CV(k_{\Delta})$ & $\beta_{\Delta}$ & $r$ & $r_{\Delta}$ & $Q$ & Comm. \\
\midrule
SIS & ER & 250.0200 & 0.2773 & 269.2406 & -0.0086 & 0.0028 & 0.1157 & 7 \\
SIS & BA12 & 85.1760 & 2.1167 & 466.7841 & -0.0317 & -0.0334 & 0.1712 & 9 \\
SIS & BA16 & 171.2760 & 1.8411 & 751.8303 & -0.0160 & -0.0259 & 0.1399 & 9 \\
SIS & BA20 & 305.0760 & 1.6425 & 1128.1007 & -0.0337 & -0.0294 & 0.1243 & 10 \\
SIS & Email & 641.7505 & 1.8253 & 2779.9879 & -0.0257 & 0.0859 & 0.4111 & 8 \\
\midrule
DW & ER & 244.4760 & 0.2900 & 265.0358 & 0.0136 & 0.0026 & 0.1111 & 11 \\
DW & BA12 & 84.2760 & 2.0504 & 438.5893 & -0.0406 & -0.0285 & 0.1665 & 9 \\
DW & Collaboration & 68.9452 & 3.9702 & 1155.7123 & 0.6392 & 0.3972 & 0.8478 & 40 \\
\midrule
Gene & ER & 252.6360 & 0.2864 & 273.3520 & 0.0028 & 0.0022 & 0.1098 & 10 \\
Gene & BA12 & 86.3760 & 2.1086 & 470.4370 & -0.0471 & -0.0315 & 0.1605 & 10 \\
Gene & Yeast & 5.0664 & 7.3651 & 279.8962 & -0.5976 & -0.0581 & 0.4525 & 15 \\
\bottomrule
\end{tabular*}
\caption{Higher-order structural, assortativity, and community-level properties of the synthetic and real networks considered for each dynamical model.}
\label{tab:structural_hoi}
\end{table}

\bibliographystyle{unsrtnat}
\bibliography{first_Draft}

\begin{thebibliography}{62}
\providecommand{\natexlab}[1]{#1}
\providecommand{\url}[1]{\texttt{#1}}
\expandafter\ifx\csname urlstyle\endcsname\relax
  \providecommand{\doi}[1]{doi: #1}\else
  \providecommand{\doi}{doi: \begingroup \urlstyle{rm}\Url}\fi

\bibitem[Holling(1973)]{holling1973resilience}
C.~S. Holling.
\newblock Resilience and stability of ecological systems.
\newblock \emph{Annual Review of Ecology and Systematics}, 4:\penalty0 1--23, 1973.
\newblock \doi{10.1146/annurev.es.04.110173.000245}.

\bibitem[Walker et~al.(2004)Walker, Holling, Carpenter, and Kinzig]{walker2004resilience}
B.~Walker, C.~S. Holling, S.~R. Carpenter, and A.~Kinzig.
\newblock Resilience, adaptability and transformability in social-ecological systems.
\newblock \emph{Ecology and Society}, 9\penalty0 (2):\penalty0 5, 2004.
\newblock \doi{10.5751/ES-00650-090205}.

\bibitem[Scheffer(2009)]{scheffer2009critical}
M.~Scheffer.
\newblock \emph{Critical Transitions in Nature and Society}.
\newblock Princeton University Press, Princeton, NJ, 2009.

\bibitem[Krakovsk{\'a} et~al.(2024)Krakovsk{\'a}, Kuehn, and Longo]{krakovska2024resilience}
H.~Krakovsk{\'a}, C.~Kuehn, and I.~P. Longo.
\newblock Resilience of dynamical systems.
\newblock \emph{European Journal of Applied Mathematics}, 35\penalty0 (1):\penalty0 155--200, 2024.
\newblock \doi{10.1017/S0956792523000141}.

\bibitem[Schoenmakers and Feudel(2021)]{schoenmakers2021resilience}
S.~Schoenmakers and U.~Feudel.
\newblock A resilience concept based on system functioning: A dynamical systems perspective.
\newblock \emph{Chaos}, 31:\penalty0 053126, 2021.
\newblock \doi{10.1063/5.0042755}.

\bibitem[Gao et~al.(2016)Gao, Barzel, and Barab{\'a}si]{gao2016universal}
J.~Gao, B.~Barzel, and A.-L. Barab{\'a}si.
\newblock Universal resilience patterns in complex networks.
\newblock \emph{Nature}, 530\penalty0 (7590):\penalty0 307--312, 2016.
\newblock \doi{10.1038/nature16948}.

\bibitem[Artime et~al.(2024)Artime, Grassia, De~Domenico, Gleeson, Makse, Mangioni, Perc, and Radicchi]{artime2024robustness}
O.~Artime, M.~Grassia, M.~De~Domenico, J.~P. Gleeson, H.~A. Makse, G.~Mangioni, M.~Perc, and F.~Radicchi.
\newblock Robustness and resilience of complex networks.
\newblock \emph{Nature Reviews Physics}, 6:\penalty0 114--131, 2024.
\newblock \doi{10.1038/s42254-023-00676-y}.

\bibitem[Liu et~al.(2022)Liu, Li, Ma, Szymanski, Stanley, and Gao]{liu2022network}
X.~Liu, D.~Li, M.~Ma, B.~B. Szymanski, H.~E. Stanley, and J.~Gao.
\newblock Network resilience.
\newblock \emph{Physics Reports}, 971:\penalty0 1--108, 2022.
\newblock \doi{10.1016/j.physrep.2022.04.002}.

\bibitem[Pikovsky et~al.(2001)Pikovsky, Rosenblum, and Kurths]{pikovsky2001synchronization}
A.~Pikovsky, M.~Rosenblum, and J.~Kurths.
\newblock \emph{Synchronization: A Universal Concept in Nonlinear Sciences}, volume~12 of \emph{Cambridge Nonlinear Science Series}.
\newblock Cambridge University Press, Cambridge, 2001.
\newblock \doi{10.1017/CBO9780511755743}.

\bibitem[Arenas et~al.(2008)Arenas, D{\'\i}az-Guilera, Kurths, Moreno, and Zhou]{arenas2008physrep}
A.~Arenas, A.~D{\'\i}az-Guilera, J.~Kurths, Y.~Moreno, and C.~Zhou.
\newblock Synchronization in complex networks.
\newblock \emph{Physics Reports}, 469:\penalty0 93--153, 2008.
\newblock \doi{10.1016/j.physrep.2008.09.002}.

\bibitem[Ji et~al.(2013)Ji, Peron, Menck, Rodrigues, and Kurths]{ji2013prl}
P.~Ji, T.~K. D.~M. Peron, P.~J. Menck, F.~A. Rodrigues, and J.~Kurths.
\newblock Cluster explosive synchronization in complex networks.
\newblock \emph{Physical Review Letters}, 110:\penalty0 218701, 2013.
\newblock \doi{10.1103/PhysRevLett.110.218701}.

\bibitem[Rodrigues et~al.(2016)Rodrigues, Peron, Ji, and Kurths]{rodrigues2016kuramoto}
F.~A. Rodrigues, T.~K. D.~M. Peron, P.~Ji, and J.~Kurths.
\newblock The kuramoto model in complex networks.
\newblock \emph{Physics Reports}, 610:\penalty0 1--98, 2016.
\newblock \doi{10.1016/j.physrep.2015.10.008}.

\bibitem[Kundu et~al.(2017)Kundu, Khanra, Hens, and Pal]{kundu2017pre}
P.~Kundu, P.~Khanra, C.~Hens, and P.~Pal.
\newblock Transition to synchrony in degree-frequency correlated sakaguchi--kuramoto model.
\newblock \emph{Physical Review E}, 96:\penalty0 052216, 2017.
\newblock \doi{10.1103/PhysRevE.96.052216}.

\bibitem[Kundu et~al.(2018)Kundu, Hens, Barzel, and Pal]{kundu2018epl}
P.~Kundu, C.~Hens, B.~Barzel, and P.~Pal.
\newblock Perfect synchronization in networks of phase-frustrated oscillators.
\newblock \emph{Europhysics Letters}, 120:\penalty0 40002, 2018.
\newblock \doi{10.1209/0295-5075/120/40002}.

\bibitem[Kundu and Pal(2019)]{kundu2019chaos}
P.~Kundu and P.~Pal.
\newblock Synchronization transition in sakaguchi--kuramoto model on complex networks with partial degree-frequency correlation.
\newblock \emph{Chaos}, 29:\penalty0 013123, 2019.
\newblock \doi{10.1063/1.5045836}.

\bibitem[Khanra et~al.(2018)Khanra, Kundu, Hens, and Pal]{khanra2018pre}
P.~Khanra, P.~Kundu, C.~Hens, and P.~Pal.
\newblock Explosive synchronization in phase-frustrated multiplex networks.
\newblock \emph{Physical Review E}, 98:\penalty0 052315, 2018.
\newblock \doi{10.1103/PhysRevE.98.052315}.

\bibitem[Khanra and Pal(2021)]{khanra2021csf}
P.~Khanra and P.~Pal.
\newblock Explosive synchronization in multilayer networks through partial adaptation.
\newblock \emph{Chaos, Solitons \& Fractals}, 143:\penalty0 110621, 2021.
\newblock \doi{10.1016/j.chaos.2020.110621}.

\bibitem[Dutta et~al.(2025{\natexlab{a}})Dutta, Pal, and Hens]{dutta2025hypergraph}
S.~Dutta, P.~Pal, and C.~Hens.
\newblock Generalized adaptation-induced non-universal synchronization transitions in random hypergraphs.
\newblock \emph{Chaos}, 35:\penalty0 113131, 2025{\natexlab{a}}.
\newblock \doi{10.1063/5.0295455}.

\bibitem[Dutta et~al.(2023{\natexlab{a}})Dutta, Kundu, Khanra, Hens, and Pal]{dutta2023perfect}
S.~Dutta, P.~Kundu, P.~Khanra, C.~Hens, and P.~Pal.
\newblock Perfect synchronization in complex networks with higher-order interactions.
\newblock \emph{Physical Review E}, 108:\penalty0 024304, 2023{\natexlab{a}}.
\newblock \doi{10.1103/PhysRevE.108.024304}.

\bibitem[Das et~al.(2025)Das, Dutta, and Pal]{das2025phaselag}
A.~B. Das, S.~Dutta, and P.~Pal.
\newblock Effect of phase-lag on synchronization in adaptive multilayer networks with higher-order interactions.
\newblock arXiv preprint arXiv:2507.01640, 2025.

\bibitem[Dutta et~al.(2023{\natexlab{b}})Dutta, Mondal, Kundu, Khanra, Pal, and Hens]{dutta2023phase}
S.~Dutta, A.~Mondal, P.~Kundu, P.~Khanra, P.~Pal, and C.~Hens.
\newblock Impact of phase lag on synchronization in frustrated kuramoto model with higher-order interactions.
\newblock \emph{Physical Review E}, 108:\penalty0 034208, 2023{\natexlab{b}}.
\newblock \doi{10.1103/PhysRevE.108.034208}.

\bibitem[Dutta et~al.(2024)Dutta, Kundu, Khanra, Hens, and Pal]{dutta2024adaptive}
S.~Dutta, P.~Kundu, P.~Khanra, C.~Hens, and P.~Pal.
\newblock Transition to synchronization in the adaptive sakaguchi--kuramoto model with higher-order interactions.
\newblock \emph{Physical Review E}, 110:\penalty0 064317, 2024.
\newblock \doi{10.1103/PhysRevE.110.064317}.

\bibitem[Dutta et~al.(2025{\natexlab{b}})Dutta, Kundu, Khanra, Minati, Boccaletti, Pal, and Hens]{dutta2025double}
S.~Dutta, P.~Kundu, P.~Khanra, L.~Minati, S.~Boccaletti, P.~Pal, and C.~Hens.
\newblock Double explosive kuramoto transition in hypergraphs.
\newblock \emph{Physical Review Research}, 7:\penalty0 L022049, 2025{\natexlab{b}}.
\newblock \doi{10.1103/PhysRevResearch.7.L022049}.

\bibitem[Ghosh et~al.(2025)Ghosh, Xue, Mishra, Saha, Dudkowski, Dana, Kapitaniak, Kurths, Ji, and Hens]{ghosh2025universal}
Subrata Ghosh, Linuo Xue, Arindam Mishra, Suman Saha, Dawid Dudkowski, Syamal~K Dana, Tomasz Kapitaniak, J{\"u}rgen Kurths, Peng Ji, and Chittaranjan Hens.
\newblock Universal nonlinear dynamics in damped and driven physical systems: From pendula via josephson junctions to power grids.
\newblock \emph{Physics Reports}, 1147:\penalty0 1--112, 2025.
\newblock \doi{https://doi.org/10.1016/j.physrep.2025.09.005}.

\bibitem[Pastor-Satorras and Vespignani(2001)]{pastor2001prl}
R.~Pastor-Satorras and A.~Vespignani.
\newblock Epidemic spreading in scale-free networks.
\newblock \emph{Physical Review Letters}, 86:\penalty0 3200--3203, 2001.
\newblock \doi{10.1103/PhysRevLett.86.3200}.

\bibitem[Granell et~al.(2013)Granell, G{\'o}mez, and Arenas]{granell2013prl}
C.~Granell, S.~G{\'o}mez, and A.~Arenas.
\newblock Dynamical interplay between awareness and epidemic spreading in multiplex networks.
\newblock \emph{Physical Review Letters}, 111:\penalty0 128701, 2013.
\newblock \doi{10.1103/PhysRevLett.111.128701}.

\bibitem[Pastor-Satorras et~al.(2015)Pastor-Satorras, Castellano, Van~Mieghem, and Vespignani]{pastor2015rmp}
R.~Pastor-Satorras, C.~Castellano, P.~Van~Mieghem, and A.~Vespignani.
\newblock Epidemic processes in complex networks.
\newblock \emph{Reviews of Modern Physics}, 87:\penalty0 925--979, 2015.
\newblock \doi{10.1103/RevModPhys.87.925}.

\bibitem[Wang et~al.(2017)Wang, Tang, Stanley, and Braunstein]{wang2017rpp}
W.~Wang, M.~Tang, H.~E. Stanley, and L.~A. Braunstein.
\newblock Unification of theoretical approaches for epidemic spreading on complex networks.
\newblock \emph{Reports on Progress in Physics}, 80:\penalty0 036603, 2017.
\newblock \doi{10.1088/1361-6633/aa5398}.

\bibitem[Mei et~al.(2017)Mei, Mohagheghi, Zampieri, and Bullo]{mei2017annualreview}
W.~Mei, S.~Mohagheghi, S.~Zampieri, and F.~Bullo.
\newblock On the dynamics of deterministic epidemic propagation over networks.
\newblock \emph{Annual Review of Control}, 44:\penalty0 116--128, 2017.
\newblock \doi{10.1016/j.arcontrol.2017.09.002}.

\bibitem[Colizza et~al.(2007)Colizza, Barrat, Barthelemy, Valleron, and Vespignani]{colizza2007bmc}
V.~Colizza, A.~Barrat, M~Barthelemy, AJ. Valleron, and A.~Vespignani.
\newblock Modeling the worldwide spread of pandemic influenza: Baseline case and containment interventions.
\newblock \emph{PLoS Med}, 4(1),e13, 2007.
\newblock \doi{10.1371/journal.pmed.0040013}.

\bibitem[Higham and de~Kergorlay(2021)]{higham2021epidemics}
D.~J. Higham and H.-L. de~Kergorlay.
\newblock Epidemics on hypergraphs: spectral thresholds for extinction.
\newblock \emph{Proceedings of the Royal Society A}, 477\penalty0 (2252):\penalty0 20210232, 2021.
\newblock \doi{10.1098/rspa.2021.0232}.

\bibitem[Yuan et~al.(2026)Yuan, Jin, and Liu]{yuan2026noise}
Rong Yuan, Zhen Jin, and Maoxing Liu.
\newblock Noise-induced transients in the propagation of epidemic with higher-order interactions.
\newblock \emph{Chaos: An Interdisciplinary Journal of Nonlinear Science}, 36\penalty0 (4):\penalty0 043101, 2026.
\newblock \doi{10.1063/5.0302152}.

\bibitem[Luo et~al.(2026)Luo, Lambiotte, and Ji]{luo2026temporal}
C.~Luo, R.~Lambiotte, and P.~Ji.
\newblock Temporal heterogeneity shapes diffusion dynamics in complex networks.
\newblock \emph{Nature Communications}, 2026.
\newblock \doi{10.1038/s41467-026-72161-w}.

\bibitem[Albert and Barab{\'a}si(2002)]{albert2002rmp}
R.~Albert and A.-L. Barab{\'a}si.
\newblock Statistical mechanics of complex networks.
\newblock \emph{Reviews of Modern Physics}, 74:\penalty0 47--97, 2002.
\newblock \doi{10.1103/RevModPhys.74.47}.

\bibitem[Barrat et~al.(2008)Barrat, Barth{\'e}lemy, and Vespignani]{barrat2008dynamical}
A.~Barrat, M.~Barth{\'e}lemy, and A.~Vespignani.
\newblock \emph{Dynamical Processes on Complex Networks}.
\newblock Cambridge University Press, Cambridge, 2008.
\newblock \doi{10.1017/CBO9780511791383}.

\bibitem[Newman(2010)]{newman2010networks}
M.~Newman.
\newblock \emph{Networks: An Introduction}.
\newblock Oxford University Press, Oxford, 2010.

\bibitem[Boccaletti et~al.(2006)Boccaletti, Latora, Moreno, Chavez, and Hwang]{boccaletti2006physrep}
S.~Boccaletti, V.~Latora, Y.~Moreno, M.~Chavez, and D.-U. Hwang.
\newblock Complex networks: Structure and dynamics.
\newblock \emph{Physics Reports}, 424:\penalty0 175--308, 2006.
\newblock \doi{10.1016/j.physrep.2005.10.009}.

\bibitem[Dorogovtsev et~al.(2008)Dorogovtsev, Goltsev, and Mendes]{dorogovtsev2008rmp}
S.~N. Dorogovtsev, A.~V. Goltsev, and J.~F.~F. Mendes.
\newblock Critical phenomena in complex networks.
\newblock \emph{Reviews of Modern Physics}, 80:\penalty0 1275--1335, 2008.
\newblock \doi{10.1103/RevModPhys.80.1275}.

\bibitem[Dawn et~al.(2026)Dawn, Meena, Rogers, and Hens]{dawn2026instability}
Subrata Dawn, Shraosi.~Ghosh, Chandrakala Meena, Tim Rogers, and Chittaranjan Hens.
\newblock Origins of instability in dynamical systems on undirected networks.
\newblock \emph{Phys. Rev. E}, 113:\penalty0 014303, Jan 2026.
\newblock \doi{10.1103/g2mp-bkhf}.
\newblock URL \url{https://link.aps.org/doi/10.1103/g2mp-bkhf}.

\bibitem[Meena et~al.(2023)Meena, Hens, Acharyya, et~al.]{Meena2023stability}
C.~Meena, C.~Hens, S.~Acharyya, et~al.
\newblock Emergent stability in complex network dynamics.
\newblock \emph{Nature Physics}, 19:\penalty0 1033--1042, 2023.
\newblock \doi{10.1038/s41567-023-02020-8}.

\bibitem[Allesina and Tang(2012)]{Allesina2012}
S.~Allesina and S.~Tang.
\newblock Stability criteria for complex ecosystems.
\newblock \emph{Nature}, 483:\penalty0 205--8, 2012.
\newblock \doi{10.1038/nature10832}.

\bibitem[Kundu et~al.(2022{\natexlab{a}})Kundu, Kori, and Masuda]{kundu2022accuracy}
P.~Kundu, H.~Kori, and N.~Masuda.
\newblock Accuracy of a one-dimensional reduction of dynamical systems on networks.
\newblock \emph{Physical Review E}, 105\penalty0 (2):\penalty0 024305, 2022{\natexlab{a}}.
\newblock \doi{10.1103/PhysRevE.105.024305}.

\bibitem[Jiang et~al.(2018)Jiang, Huang, Seager, Lin, Grebogi, Hastings, and Lai]{jiang2018predicting}
Junjie Jiang, Zonghua Huang, Thomas~P. Seager, Weigang Lin, Celso Grebogi, Alan Hastings, and Ying-Cheng Lai.
\newblock Predicting tipping points in mutualistic networks through dimension reduction.
\newblock \emph{Proceedings of the National Academy of Sciences}, 115\penalty0 (4):\penalty0 E639--E647, 2018.
\newblock \doi{10.1073/pnas.1714958115}.

\bibitem[Laurence et~al.(2019)Laurence, Doyon, Dub{\'e}, and Desrosiers]{laurence2019prx}
E.~Laurence, N.~Doyon, L.~J. Dub{\'e}, and P.~Desrosiers.
\newblock Spectral dimension reduction of complex dynamical networks.
\newblock \emph{Physical Review X}, 9:\penalty0 011042, 2019.
\newblock \doi{10.1103/PhysRevX.9.011042}.

\bibitem[MacLaren et~al.(2023)MacLaren, Kundu, and Masuda]{MaclarenJROS2023}
Neil~G. MacLaren, Prosenjit Kundu, and Naoki Masuda.
\newblock Early warnings for multi-stage transitions in dynamics on networks.
\newblock \emph{Journal of The Royal Society Interface}, 20\penalty0 (200):\penalty0 20220743, 03 2023.
\newblock ISSN 1742-5689.
\newblock \doi{10.1098/rsif.2022.0743}.
\newblock URL \url{https://doi.org/10.1098/rsif.2022.0743}.

\bibitem[Thibeault et~al.(2020)Thibeault, St-Onge, Dub{\'e}, and Desrosiers]{thibeault2020threefold}
V.~Thibeault, G.~St-Onge, L.~J. Dub{\'e}, and P.~Desrosiers.
\newblock Threefold way to the dimension reduction of dynamics on networks: An application to synchronization.
\newblock \emph{Physical Review Research}, 2:\penalty0 043215, 2020.
\newblock \doi{10.1103/PhysRevResearch.2.043215}.

\bibitem[Burgio et~al.(2021)Burgio, Arenas, G{\'o}mez, and Matamalas]{burgio2021compphys}
G.~Burgio, A.~Arenas, S.~G{\'o}mez, and J.~T. Matamalas.
\newblock Network clique cover approximation to analyze complex contagions through group interactions.
\newblock \emph{Communications Physics}, 4:\penalty0 111, 2021.
\newblock \doi{10.1038/s42005-021-00618-z}.

\bibitem[Battiston et~al.(2021)Battiston, Amico, Barrat, Bianconi, Ferraz~de Arruda, Franceschiello, Iacopini, K{\'e}fi, Latora, Moreno, Musciotto, Nicosia, Peixoto, Rosvall, Schaub, Sciarra, Vaccarino, and Petri]{battiston2021natphys}
F.~Battiston, E.~Amico, A.~Barrat, G.~Bianconi, G.~Ferraz~de Arruda, B.~Franceschiello, I.~Iacopini, S.~K{\'e}fi, V.~Latora, Y.~Moreno, F.~Musciotto, V.~Nicosia, T.~P. Peixoto, M.~Rosvall, M.~T. Schaub, C.~Sciarra, F.~Vaccarino, and G.~Petri.
\newblock The physics of higher-order interactions in complex systems.
\newblock \emph{Nature Physics}, 17:\penalty0 1093--1098, 2021.
\newblock \doi{10.1038/s41567-021-01371-4}.

\bibitem[Boccaletti et~al.(2023{\natexlab{a}})Boccaletti, De~Lellis, del Genio, Alfaro-Bittner, Criado, Jalan, and Romance]{boccaletti2023structure}
S.~Boccaletti, P.~De~Lellis, C.~I. del Genio, K.~Alfaro-Bittner, R.~Criado, S.~Jalan, and M.~Romance.
\newblock The structure and dynamics of networks with higher order interactions.
\newblock \emph{Physics Reports}, 1018:\penalty0 1--64, 2023{\natexlab{a}}.
\newblock \doi{10.1016/j.physrep.2023.04.002}.

\bibitem[Bick et~al.(2023)Bick, Gross, Harrington, and Schaub]{bick2023higher}
C.~Bick, T.~Gross, H.~A. Harrington, and M.~T. Schaub.
\newblock What are higher-order networks?
\newblock \emph{SIAM Review}, 65:\penalty0 686--731, 2023.
\newblock \doi{10.1137/21M1414024}.

\bibitem[Boccaletti et~al.(2023{\natexlab{b}})Boccaletti, {De Lellis}, {del Genio}, Alfaro-Bittner, Criado, Jalan, and Romance]{ji2023physrep}
S.~Boccaletti, P.~{De Lellis}, C.I. {del Genio}, K.~Alfaro-Bittner, R.~Criado, S.~Jalan, and M.~Romance.
\newblock The structure and dynamics of networks with higher order interactions.
\newblock \emph{Physics Reports}, 1018:\penalty0 1--64, 2023{\natexlab{b}}.
\newblock \doi{10.1016/j.physrep.2023.04.002}.

\bibitem[von~der Gracht et~al.(2024)von~der Gracht, Nijholt, and Rink]{Gracht2024}
Sören von~der Gracht, Eddie Nijholt, and Bob Rink.
\newblock Higher-order interactions lead to ‘reluctant’ synchrony breaking.
\newblock \emph{Proceedings of the Royal Society A: Mathematical, Physical and Engineering Sciences}, 480\penalty0 (2301):\penalty0 20230945, 11 2024.
\newblock ISSN 1364-5021.
\newblock \doi{10.1098/rspa.2023.0945}.
\newblock URL \url{https://doi.org/10.1098/rspa.2023.0945}.

\bibitem[Majhi et~al.(2022)Majhi, Perc, and Ghosh]{Majhi2022}
Soumen Majhi, Matjaž Perc, and Dibakar Ghosh.
\newblock Dynamics on higher-order networks: a review.
\newblock \emph{Journal of The Royal Society Interface}, 19\penalty0 (188):\penalty0 20220043, 03 2022.
\newblock ISSN 1742-5689.
\newblock \doi{10.1098/rsif.2022.0043}.
\newblock URL \url{https://doi.org/10.1098/rsif.2022.0043}.

\bibitem[Iacopini et~al.(2019)Iacopini, Petri, Barrat, and Latora]{iacopini2019simplicial}
I.~Iacopini, G.~Petri, A.~Barrat, and V.~Latora.
\newblock Simplicial models of social contagion.
\newblock \emph{Nature Communications}, 10:\penalty0 2485, 2019.
\newblock \doi{10.1038/s41467-019-10431-6}.

\bibitem[Ferraz~de Arruda et~al.(2020)Ferraz~de Arruda, Petri, and Moreno]{arruda2020contagion}
G.~Ferraz~de Arruda, G.~Petri, and Y.~Moreno.
\newblock Social contagion models on hypergraphs.
\newblock \emph{Physical Review Research}, 2:\penalty0 023032, 2020.
\newblock \doi{10.1103/PhysRevResearch.2.023032}.

\bibitem[Grilli et~al.(2017)Grilli, Barab{\'a}s, Michalska-Smith, and Allesina]{grilli2017higher}
J.~Grilli, G.~Barab{\'a}s, M.~J. Michalska-Smith, and S.~Allesina.
\newblock Higher-order interactions stabilize dynamics in competitive network models.
\newblock \emph{Nature}, 548:\penalty0 210--213, 2017.
\newblock \doi{10.1038/nature23273}.

\bibitem[Sun and Bianconi(2021)]{sun2021hypergraph}
F.~Sun and G.~Bianconi.
\newblock Higher-order percolation processes on multiplex hypergraphs.
\newblock \emph{Physical Review E}, 104:\penalty0 034306, 2021.
\newblock \doi{10.1103/PhysRevE.104.034306}.

\bibitem[Kachhvah and Jalan(2022)]{kachhvah2022simplicial}
A.~D. Kachhvah and S.~Jalan.
\newblock Hebbian plasticity rules abrupt desynchronization in pure simplicial complexes.
\newblock \emph{New Journal of Physics}, 24:\penalty0 052002, 2022.
\newblock \doi{10.1088/1367-2630/ac6bba}.

\bibitem[Ghosh et~al.(2023)Ghosh, Khanra, Kundu, Ji, Ghosh, and Hens]{ghosh2023chaos}
S.~Ghosh, P.~Khanra, P.~Kundu, P.~Ji, D.~Ghosh, and C.~Hens.
\newblock Dimension reduction in higher-order contagious phenomena.
\newblock \emph{Chaos}, 33:\penalty0 053117, 2023.
\newblock \doi{10.1063/5.0152959}.

\bibitem[Wang et~al.(2026)Wang, Zhu, and Liu]{wang2026network}
Zheng Wang, Jinjie Zhu, and Xianbin Liu.
\newblock Network stochastic resonance under higher-order interactions.
\newblock \emph{Proceedings of the Royal Society A: Mathematical, Physical and Engineering Sciences}, 482\penalty0 (2333):\penalty0 20250945, 2026.
\newblock \doi{10.1098/rspa.2025.0945}.
\newblock URL \url{https://doi.org/10.1098/rspa.2025.0945}.

\bibitem[Kundu et~al.(2022{\natexlab{b}})Kundu, MacLaren, Kori, and Masuda]{kundu2022rspa}
P.~Kundu, N.~MacLaren, H.~Kori, and N.~Masuda.
\newblock Mean-field theory for double-well systems on degree-heterogeneous networks.
\newblock \emph{Proceedings of the Royal Society A}, 478:\penalty0 20220350, 2022{\natexlab{b}}.
\newblock \doi{10.1098/rspa.2022.0350}.

\bibitem[Landry and Restrepo(2022)]{landry2022hypergraph}
N.~W. Landry and J.~G. Restrepo.
\newblock Hypergraph assortativity: A dynamical systems perspective.
\newblock \emph{Chaos: An Interdisciplinary Journal of Nonlinear Science}, 32\penalty0 (5):\penalty0 053113, 2022.
\newblock \doi{10.1063/5.0086905}.

\end{thebibliography}

\end{document}